\documentclass[A_4paper,11pt, final]{article}
\pdfoutput=1
\usepackage[utf8]{inputenc}
\usepackage{jheppub}
\usepackage{amsmath}
\usepackage{float}
\usepackage{graphicx}
\usepackage{subcaption}
\usepackage{braket}

\DeclareMathOperator{\arcsinh}{arcsinh}
\DeclareMathOperator{\arccosh}{arccosh}

\usepackage[inline]{showlabels}

\graphicspath{{./figs/}}

\newcommand*\diff{\mathrm{d}}

\usepackage{xspace}
\newcommand*{\ie}{i.e., }
\newcommand*{\eg}{e.g., }

\newcommand*{\eq}{eq.\@\xspace}
\newcommand*{\eqs}{eqs.\@\xspace}
\newcommand*{\cf}{cf.\@\xspace}

\title{Critical Points in Palatini Higgs Inflation with Small Non-Minimal Coupling}

\author{Arthur Poisson,}
\author[a]{Inar Timiryasov,}
\author[b]{Sebastian Zell}

\affiliation[a]{Niels Bohr Institute, University of Copenhagen, Blegdamsvej 17, DK-2010, Copenhagen, Denmark}
\affiliation[b]{Centre for Cosmology, Particle Physics and Phenomenology -- CP3,
	Universit\'e catholique de Louvain, B-1348 Louvain-la-Neuve, Belgium}

\emailAdd{arthurpoisson99@icloud.com}
\emailAdd{inar.timiryasov@nbi.ku.dk}
\emailAdd{sebastian.zell@uclouvain.be}

\abstract{We investigate inflation driven by the Higgs boson in the Palatini formulation of General Relativity.
Our analysis primarily focuses on a small non-minimal coupling of the Higgs field to gravity in the range $0<\xi\lesssim1$. We incorporate the renormalization group running of the relevant parameters as computed within the Standard Model and allow for small corrections.
In addition to $\xi$, our model features two tunable parameters: the low-energy value of the top Yukawa coupling and an effective jump of the Higgs self-interaction. Our results indicate that critical points leading to a large enhancement of the power spectrum can be produced. However,  the observed amplitude of perturbations in the CMB cannot be matched within this setting. On the one hand, this makes it difficult to generate a sizable abundance of primordial black holes. On the other hand, our finding can be viewed as further evidence that Palatini Higgs inflation has favourable high-energy properties due to robustness against quantum corrections.}

\date{\today}

\makeatletter
\gdef\@fpheader{\phantom{text}}
\makeatother

\begin{document}
	\maketitle
	
	\section{Introduction}
	\label{sec:intro}

There are strong indications that a phase of accelerated expansion occurred in the first moments of our Universe. This proposed concept of inflation can account for the observed homogeneity and isotropy on macroscopic scales. Furthermore, it provides an intriguing explanation for small initial inhomogeneities arising from quantum fluctuations \cite{Starobinsky:1980te, Guth:1980zm, Linde:1981mu, Mukhanov:1981xt}. Since a plethora of inflationary models have been developed (see \cite{Martin:2013tda}), a significant challenge is selecting the one that nature has realized.

In this work, we consider the possibility that inflation was caused by the Higgs boson of the Standard Model (SM) \cite{Bezrukov:2007ep}. This proposal, known as Higgs Inflation (HI), is supported by two classes of observations. First, it is one of the inflationary scenarios that aligns excellently with current measurements of the cosmic microwave background (CMB) \cite{Planck:2018jri, BICEP:2021xfz}. Second, HI is the only model that can be realized exclusively with particles that have already been detected in experiments, thereby relating it to the most recent breakthrough discovery in particle physics \cite{ATLAS:2012yve,CMS:2012qbp}. The only required ingredient is a non-minimal coupling $\xi$ between the Higgs field and the gravitational Ricci scalar. Its numerical value is chosen according to the observed amplitude of perturbations in the CMB, which leads to $\xi \gg 1$.

However, there exists more than one version of HI, which is due to the fact that General Relativity (GR) exists in different formulations (see \cite{Rigouzzo:2022yan} for an overview). All of them are fully equivalent in pure gravity and are therefore considered as different incarnations of one and the same theory of GR. Once matter is included, however, the various versions of GR lead to diverse phenomenology. HI is a prime example where observable predictions depend sensitively on the chosen formulation of GR. This became apparent when comparing the metric \cite{Bezrukov:2007ep} and Palatini \cite{Bauer:2008zj} versions and has been subsequently investigated in several other formulations of GR \cite{Rasanen:2018ihz, Raatikainen:2019qey, Langvik:2020nrs, Shaposhnikov:2020gts,Rigouzzo:2022yan} (also see \cite{Azri:2017uor, Jarv:2021ehj}).\footnote
{Reviews of metric and Palatini Higgs inflation are provided in \cite{Rubio:2018ogq} and \cite{Tenkanen:2020dge}, respectively.
}

At first glance, it seems that HI can be directly related to the properties of the Higgs boson as measured in collider experiments. However, the existence of such a connection significantly depends on radiative corrections, which were initially investigated in the metric version of HI \cite{Barvinsky:2008ia,Bezrukov:2008ej,DeSimone:2008ei,Burgess:2009ea,Barbon:2009ya,Bezrukov:2009db, Barvinsky:2009fy, Barvinsky:2009ii, Bezrukov:2010jz,Prokopec:2014iya,Hamada:2014iga,Bezrukov:2014bra,Ren:2014sya,Bezrukov:2014ipa, Fumagalli:2016lls, Enckell:2016xse,Escriva:2016cwl,Bezrukov:2017dyv,Fumagalli:2017cdo,Enckell:2018kkc,Antoniadis:2021axu, Mikura:2021clt, Ito:2021ssc,Karananas:2022byw}. One notable effect is that the values of coupling constants at inflationary scales generically differ from their low-energy counterparts due to the renormalization group (RG) running \cite{Bezrukov:2007ep,DeSimone:2008ei,Bezrukov:2008ej,Bezrukov:2009db,Barvinsky:2009fy,Barvinsky:2009ii,Hamada:2014iga,Bezrukov:2014bra,Bezrukov:2014ipa,Fumagalli:2016lls,Enckell:2016xse,Bezrukov:2017dyv,Fumagalli:2017cdo,Enckell:2018kkc}. More dramatically, a cutoff energy scale $\Lambda$ emerges, above which perturbation theory breaks down \cite{Burgess:2009ea,Barbon:2009ya}.\footnote
{There are explicit UV-completions for metric Higgs inflation \cite{Giudice:2010ka,Gorbunov:2013dqa,Barbon:2015fla,Ema:2017rqn} while no such model exists in the Palatini case \cite{Mikura:2021clt}.}
The low value of $\Lambda$ in the metric scenario generically prevents a direct connection between collider measurements and inflationary physics \cite{Bezrukov:2010jz,Bezrukov:2014ipa}. This phenomenon can be understood in terms of effective jumps of coupling constants, which occur when moving from low to high energies, the values of which are a priori undetermined \cite{Bezrukov:2014ipa}. Moreover, $\Lambda$ renders an unambiguous computation of reheating, i.e., the transition from inflation to a radiation-dominated Universe, impossible \cite{Ema:2016dny} (see also \cite{He:2018mgb,Bezrukov:2020txg}). 

The Palatini version of HI presents more favourable properties due to a higher cutoff scale \cite{Bauer:2010jg}. Nonetheless, subsequent studies of quantum effects are less numerous \cite{Rasanen:2017ivk, Markkanen:2017tun,Enckell:2018kkc,Rasanen:2018fom,Jinno:2019und,Shaposhnikov:2020fdv,Enckell:2020lvn,Antoniadis:2021axu,Mikura:2021clt,Ito:2021ssc,Karananas:2022byw}.\footnote
 {Loop corrections are expected to be small \cite{Markkanen:2017tun} whereas including generic additional contributions, which do not respect the asymptotic shift symmetry of the Einstein frame potential, can easily change predictions or make inflation impossible \cite{Jarv:2017azx,Jinno:2019und}.}
As a result, the effects of RG running are calculable within the SM \cite{Shaposhnikov:2020fdv}, unless contributions due to unknown physics at high energies are unnaturally large. Correspondingly, the effective jumps of coupling constants are small, if present at all \cite{Shaposhnikov:2020fdv}. This allows for a connection between measurements at collider experiments and inflationary observables. Specifically, successful Palatini HI implies an upper bound on the top Yukawa coupling as measured at colliders \cite{Shaposhnikov:2020fdv}. Moreover, reheating can be unambiguously computed \cite{Rubio:2019ypq,Dux:2022kuk}.

Quantum corrections can cause critical points in an inflationary potential, such as an inflection point or a small local minimum. These lead to an enhancement of the generated power spectrum. If it is sufficiently large, primordial black holes (PBHs) are produced \cite{Zeldovich:1967lct,Hawking:1971ei,Carr:1974nx,Chapline:1975ojl} (see \cite{Carr:2020xqk,Green:2020jor,Escriva:2022duf} for recent reviews), although determining the resulting abundance of PBHs precisely remains a challenge \cite{Franciolini:2018vbk,Ezquiaga:2018gbw,Cruces:2018cvq,Ezquiaga:2019ftu,Figueroa:2020jkf,Figueroa:2021zah,Hooshangi:2021ubn,Cai:2021zsp,Animali:2022otk,Cruces:2022dom,Mishra:2023lhe,Kawaguchi:2023mgk,Hooshangi:2023kss,Stamou:2023vxu}. Moreover, a question has emerged regarding whether a sizable amount of PBHs can be produced while retaining the validity of inflationary perturbation theory \cite{Kristiano:2022maq,Riotto:2023hoz,Choudhury:2023vuj,Kristiano:2023scm,Riotto:2023gpm,Choudhury:2023jlt,Choudhury:2023rks,Firouzjahi:2023aum,Firouzjahi:2023ahg,Franciolini:2023lgy}.

There are various proposals for producing PBHs from metric HI \cite{Ezquiaga:2017fvi,Rasanen:2018fom,Drees:2019xpp,Yi:2020cut,Yi:2021lxc}, where different assumptions and models for radiative corrections were employed (see also \cite{Garcia-Bellido:2017mdw,Motohashi:2017kbs,Germani:2017bcs,Ballesteros:2017fsr,Ballesteros:2020qam,Karam:2022nym,Bhatt:2022mmn} for further discussions about PBHs from single-field inflation and \cite{Ozsoy:2023ryl} for a review and additional references).\footnote
{PBHs can also be generated in UV-completions of metric HI \cite{Cheong:2019vzl,Gundhi:2020zvb,Cheong:2022gfc}.}
These works typically feature three or four tunable parameters. The latter number generically allows to match the observed amplitude of perturbations in the CMB and to simultaneously tune the height, width and location of a feature in the inflationary potential.

 An investigation of the generation of PBHs from Palatini HI was initiated in \cite{Rasanen:2018fom}, where effective jumps in the coupling constants were taken into account in addition to the effects of RG running as computed within the SM. For appropriately chosen and sufficiently large values of the jumps, a critical point in the inflationary potential leads to the generation of a sizable abundance of PBHs \cite{Rasanen:2018fom}. The effects of radiative corrections reduce the value of $\xi$ as compared to the tree-level result but still $\xi\gg 1$. A subsequent study, which focus on features in the inflationary potential of Palatini HI, confirmed these findings \cite{Enckell:2020lvn} (see also \cite{Figueroa:2020jkf,Figueroa:2021zah} for related works in metric HI).
 
 In the present work, our goal is to determine if critical point in Palatini HI can also be generated with slightly stronger assumptions about radiative effects; namely, we shall only include small effective jumps as corrections to RG running as computed within the SM. To this end, we focus on a feature that is present already in the SM: It is known that the running of the Higgs self-coupling $\lambda$ has a local minimum slightly below $M_P$ (see \eg \cite{Bezrukov:2014ina}). Therefore, we choose parameters such that inflation takes place near this feature. Since the energy scale of Palatini HI is parametrically given as $M_P/\sqrt{\xi}$, this is only possible for $0 \lesssim\xi\lesssim 1$. Such values are additionally interesting because then the action does not feature any large ($\gg 1$) coupling constants. Whereas at tree level $\xi \gg 1$ is required in order to match the observed amplitude of perturbations in the CMB, this is not necessarily true any more once effects of RG running with corresponding tunable parameters are taken into account.

The paper is structured as follows. We start in section \ref{sec:GenInfl} with a review of known results related to inflation. In section \ref{sec:HiggsInflation}, we provide the theoretical background of Higgs Inflation within the framework of Palatini General Relativity. We outline the theory and proceed to discuss tree-level inflation. RG Running within the Standard Model and the relevance of effective jumps are also addressed. 
Section \ref{ParamSpa} is dedicated to an analysis of the parameter space that might accommodate a large enhancement of perturbations. Within this context, section \ref{ssec:jumpless} and \ref{ssec:withJump} present a comparative study of scenarios without and with a jump, respectively. Section \ref{ssec:jumpParameterSpace} features a numerical scan over the $(\xi, \delta\lambda)$ space. 
Section \ref{sec:CMB} discusses the issue of compatibility with CMB normalization. Our numerical analysis is presented in section \ref{sec:Numerics}. 
We summarise the main findings of our study in section \ref{sec:conclusion}. 
Finally, appendix \ref{app:SmallXi} addresses the limit of a small non-minimal coupling.

\textbf{Remark}. Preliminary results of the present investigation already appeared in the master thesis by one of us \cite{Poisson:2022}.

\textbf{Conventions}. We work in natural units $M_P=\hbar=c=1$, where $M_P$ is the reduced Planck mass, and use the metric signature $(-1,+1,+1,+1)$.  

\section{Inflation and Cosmological Perturbations}\label{sec:GenInfl}

As stated in the introduction, cosmic inflation  \cite{Starobinsky:1980te, Guth:1980zm, Linde:1981mu, Mukhanov:1981xt} provides a possible origin for the observed density fluctuations in the large scale structures of our Universe. During an inflationary phase of accelerated expansion, the vacuum fluctuation of quantum field are stretch and become the seeds for primordial inhomogeneities.
To appropriately describe this mechanism, the standard approach is semi-classical, where quantum fluctuations are considered on top of a classical background solution. Our discussion will then be split in three parts: the classical evolution, the quantized first order perturbations and the mechanism we considered to generate a large enhancement of the latter.

\subsection{Background Physics}
On large scales, our Universe can be approximated as spatially homogeneous, isotropic and flat. The unique possible metric corresponding to this geometry is that of a spatially flat FLRW universe \cite{Baumann:2022mni,Gorbunov:2011zzc}:
	\begin{equation}\label{FLRW}
		\diff s^2 = - \diff t^2 + a^2(t)\diff \textbf{x}^2 \,,
	\end{equation}
where $a(t)$ is the scale factor.
	In the simple single-field realization of inflation, the total energy density is assumed to be dominated by a classical homogeneous scalar field $\chi(t)$ with a potential $U(\chi)$:
	\begin{equation}\label{ActBackField}
		S = \int \diff^4 x \sqrt{-g}\left(\frac{(\Dot{\chi})^2}{2} - U(\chi)\right) \,.
	\end{equation}
	The dynamics of such a simple system is given by the combination of the Friedman equations,
	\begin{equation} \label{Friedman}
		H^2  = \frac{1}{3}\left(\frac{\Dot{\chi}^2}{2} + U(\chi)\right) \,, \qquad 	\epsilon_H = \frac{3\Dot{\chi}^2}{\Dot{\chi}^2 + 2 U(\chi)} \,,
	\end{equation}
	 and the Klein-Gordon equation
	 \begin{equation}
		\ddot{\chi} + 3 H\Dot{\chi} + U'(\chi) = 0 \,,
	\end{equation}
	where we defined the Hubble expansion rate $H$ and what is called the \textit{first geometrical slow roll parameter} $\epsilon_H$ as \cite{Baumann:2022mni,Gorbunov:2013dqa,Ballesteros:2017fsr,Riotto:2018pcx}
	\begin{equation}\label{defHdefEps}
		H = \frac{\Dot{a}}{a}\,,\qquad\epsilon_H = -\frac{\Dot{H}}{H^2} \,.
	\end{equation}
	Physically, $\epsilon_H$ quantifies how much the universe of interest is a deformation of a de-Sitter space-time where $H$ would be a constant \cite{Baumann:2022mni,Hawking:1973uf}.
	
	From its definition \eqref{defHdefEps}, $\epsilon_H$ can be related to the acceleration of expansion as:
	\begin{equation}
		\frac{\ddot{a}}{a} = H^2(1-\epsilon_H)\;.
	\end{equation}
	Hence, the acceleration would be positive as long as $\epsilon_H<1$. Using the second Friedman equation in \eqref{Friedman}, this can be translated to a condition on the field phase space:
	\begin{equation}\label{CondInf}
		\epsilon_H<1\iff \Dot{\chi}^2 < U(\chi) \,.
	\end{equation}
	In order to solve the system analytically, one can make this condition extreme and the system is then evolving according to the so-called slow roll (SR) regime \cite{Mukhanov:1981xt}:
	\begin{equation}\label{SR1}
		\epsilon_H\ll 1 \iff \Dot{\chi}^2\ll U(\chi) \,.
	\end{equation}
	To ensure us that the SR dynamics will be stable in time, one typically assumes the variation of $\epsilon_H$ to be small:
	\begin{equation}\label{SR2}
		\frac{\Dot{\epsilon_H}}{H \epsilon_H} \sim 2\frac{\ddot{\chi}}{H\Dot{\chi}} + \mathcal{O}(\epsilon_H)\ll 1 \,.
	\end{equation}
	A nice way to consistently impose these two conditions simultaneously is to introduce a \textit{second geometric slow roll parameter} defined as:
	\begin{equation} \label{etaEpsilon}
		\eta_H = \epsilon_H - \frac{1}{2 H}\frac{\Dot{\epsilon_H}}{\epsilon_H} \,.
	\end{equation}
	The two conditions \eqref{SR1} and \eqref{SR2} are then equivalent to require that both of these parameters are small.
	
	In this regime, the relevant system of equations becomes:
	\begin{align}
			3 H^2 & =  U(\chi)  \,,\\
			\Dot{\chi}^2 & =  \frac{(U'(\chi))^2}{3 U(\chi)} \,.
	\end{align}
	Within the approximation, the geometric SR parameters coincide with some well-chosen combination of derivative of the potential:
	\begin{equation}\label{EpsEtaSR}
		\epsilon_H\sim \epsilon^{S R} = \frac{1}{2}\left(\frac{U'(\chi)}{U(\chi)}\right)^2,\quad \eta_H \sim \eta^{SR} = \frac{U''(\chi)}{U(\chi)}\;.
	\end{equation}
	In the analytical treatment of SR, inflation ends through an $\mathcal{O}(1)$ violation of the conditions \eqref{SR1} or \eqref{SR2} but with the SR parameters instead of the geometric ones. The exact moment of the end of inflation is a matter of definitions and for concreteness, we shall use the term inflation for a positive sign in the acceleration of the scale factor in order to allow the physical lengths to grow faster than the Hubble scale. This corresponds to setting $\epsilon_H = 1$. We remark, however, that $\eta_H=1$ may sometimes happen slightly before and be more appropriate.

	\subsection{First Order Perturbation Theory}
	Now that we know how to describe the full classical background dynamics, one has to deal with the quantum fluctuation around it:
	\begin{equation}
		g_{\mu\nu} = g^{FLRW}_{\mu\nu} + h_{\mu\nu} \,, \qquad \chi(\textbf{x},t) = \chi_{cl}(t) + \delta\chi(\textbf{x},t) \,.
	\end{equation}
	Before quantizing the theory, we must be a bit careful about which are the degrees of freedom in this theory. Indeed, the metric perturbation being a massless spin $2$, it can only account for $2$ degrees of freedom while the inflaton scalar counts as $1$ \cite{Weinberg:1995mt}.
	Moreover, it can be seen from the linearized Einstein equation that perturbations of different spins does not interact at the first order in perturbation theory. Therefore, our theory should be carrying only $1$ scalar degree of freedom \cite{Baumann:2022mni,Gorbunov:2013dqa,Riotto:2018pcx}.
	Choosing the flat gauge we can put all the relevant physical information in the inflaton fluctuation $\delta\chi$. 
	The result for the quadratic action is then:
	\begin{equation}\label{QuadActPert}
		S_{(2)} =\frac{1}{2} \int \diff^3 x \diff\tau\left\{\left(\frac{\diff f}{\diff\tau}\right)^2-(\nabla f)^2 + \frac{1}{z}\frac{\diff^2 z}{\diff\tau^2}f^2\right\} \,,
	\end{equation}
	where $\diff\tau = a \diff t$ is so-called conformal time, $f$ is the Mukhanov-Sasaki variable \cite{Mukhanov:1981xt}
	\begin{equation}
		f = a\delta\chi \,,
	\end{equation}
	and $z$ is a quantity including the background dependence
	\begin{equation}
		z = a\frac{\Dot{\chi}_{cl}}{H} = a\sqrt{2\epsilon_H} \,.
	\end{equation}
	The corresponding equation of motion can be obtained by varying the action \eqref{QuadActPert} with respect to $f$ leaving us with the so-called Mukhanov-Sasaki equation:
	\begin{equation}\label{MukhanovConf}
		\frac{\diff^2 f_{\textbf{k}}}{\diff\tau^2} + \left( k^2 -  \frac{1}{z}\frac{\diff^2 z}{\diff\tau^2}\right) f_{\textbf{k}}^2 = 0 \,.
	\end{equation}
	This can be equivalently obtained by direct linearization of Einstein equation \cite{Baumann:2022mni,Gorbunov:2013dqa,Riotto:2018pcx}.
	
	The action \eqref{QuadActPert} can be seen as a free massive particle with a time-depending mass. Thus, the field can be expressed within the Heisenberg picture in terms of annihilation and creation operators where the mode function respecting equation \eqref{MukhanovConf}:
	\begin{equation}
	\hat{f}_\textbf{k}(\tau) = f_k(\tau)\hat{a}_{\textbf{k}}(\tau_i)e^{i k\tau} + f_k^\star(\tau)\hat{a}^\dagger_{-\textbf{k}}(\tau_i)e^{-i k\tau} \,.
	\end{equation}
	As it is usually the case in quantum field theory in curved space-time, the only tricky point of this procedure is the non-uniqueness of the vacuum state. Indeed, being in Heisenberg picture forbids the states to evolve and so does the vacuum. The latter is then defined as the states which is annihilated by the ladder operator $\hat{a}$ evaluated at some arbitrary time $\tau_i$. The most common choice is to pick the formal limit when $\tau_i\to -\infty$, \ie when the mode of interest is still well inside the Hubble radius. Together with the canonical commutation relation, which must be respected at each time, this provides us an initial condition for the mode function $f_{k}$ known as the Bunch-Davis vacuum:
	\begin{equation}\label{BunchDavis}
	\lim_{k\tau \to -\infty}	f_{k}(\tau) = \frac{1}{\sqrt{2 k}}e^{-i k \tau}.
	\end{equation}

	Now, let us discuss the observable output of perturbation theory. At the end of inflation, the inflaton field is not a suitable variable any more. Indeed, a more natural choice would be the scalar curvature perturbation which can be translated into the initial density fluctuation of the hot Big Bang. Thus, one has to change the gauge in order to put all the relevant information in the gravity sector.
	The natural variable is then what is called the comoving curvature perturbation $\mathcal{R}$ which corresponds to the perturbation of the spatial curvature fluctuation in the gauge where the inflaton does not fluctuate any more.
	The Mukhanov-Sasaki variable can be written in this gauge as:
	\begin{equation}
		f_{k} = - z\mathcal{R}_k \,.
	\end{equation}
	The natural observable will be the late time correlators of $\mathcal{R}$ and the leading contribution is then given by the late time two-point function:
	\begin{equation}
		\braket{\mathcal{R}^2(\tau\to 0)} = \int \diff\ln k \mathcal{P}_\mathcal{R} = \int \diff\ln k \frac{k^3}{2\pi^2}\left|\frac{f_{k}(\tau\to 0)}{z}\right|^2 \,,
	\end{equation}
	where we defined the power spectrum by:
	\begin{equation}
		\mathcal{P}_\mathcal{R} = \frac{k^3}{2\pi^2}\left|\frac{f_{k}}{z}\right|^2_{k\ll a H} \,.
	\end{equation}
	Within the SR approximation, \eq \eqref{MukhanovConf} can be exactly solved, and we can give an explicit expression for the power spectrum:
	\begin{equation}\label{PowSR}
		\mathcal{P}^{SR}_\mathcal{R} = \frac{1}{8\pi^2}\left(\frac{H^2}{\epsilon_H}\right)_{k=a H}\sim \frac{1}{24\pi^2}\left(\frac{U}{\epsilon^{SR}}\right)\,.
	\end{equation}
	If our model of inflation starts within the SR dynamics, the second part of \eq \eqref{PowSR} can be used to constrain the shape of the initial shape of the potential.
	
	Indeed, amplitude of the primordial power spectrum as observed in the CMB is \cite{Planck:2018jri}:
	\begin{equation}\label{PLANCK}
		\mathcal{P}_{CMB} \sim 2.1\cdot 10^{-9}\iff \frac{U_{CMB}}{\epsilon^{SR}_{CMB}} \sim 5\cdot 10^{-7}.
	\end{equation}
	In what follows, we will refer at these two constraints as \textit{CMB normalization}.
	
	To be complete, let us also mentioned two important observables of SR inflationary models. In the derivation above, we did not mention the neither the possible small deviations of \eqref{PowSR} from an exact scale invariance not the possibility to generate tensor modes during inflation.
These two things are respectively quantified by the spectral index $n_s$ and the tensor to scalar ratio $r$ which are defined as \cite{Baumann:2022mni,Riotto:2018pcx,Gorbunov:2013dqa}:
\begin{equation}
	n_s -1= \frac{\diff \mathcal{P}_\mathcal{R}}{\diff \log k} =  2\eta^{S R}-6\epsilon^{S R}\;,
\end{equation}
and
\begin{equation}
	r = \left(\frac{\mathcal{P}_\mathcal{R}}{\mathcal{P}_{T}}\right)_{k = a H} = 16\epsilon^{SR}\;.
\end{equation}
The values of these observables at the CMB generation are also strongly constrained by the PLANCK data~\cite{Planck:2018jri} as:
\begin{equation}\label{nsPLANCK}
	n_s = 0.9665 \pm 0.0038\;,
\end{equation}
and
\begin{equation}\label{rPLANCK}
	r < 0.056\;.
\end{equation}
	
\subsection{Enhancement of the power spectrum: Ultra Slow Roll Phase} \label{ssec:USR}
Let us see the conditions under which a large enhancement of the scalar power spectrum can be produced from a model of single field inflation fully characterized by the functional form of its Einstein frame potential. 
In what follows, we will use exclusively the number of e-fold $\diff N = H \diff t$ for convenience. (The dynamical equations \eqref{Friedman} and \eqref{MukhanovConf} in this variable can be found in section \ref{sec:Numerics}.) 
	The first guess would be to look at the explicit form of the curvature power spectrum in the slow-roll approximation:
	\begin{equation}
		\mathcal{P}_\mathcal{R} = \frac{H^2}{8\pi^2\epsilon^{SR}} \,.
		\label{PSR}
	\end{equation}
	By definition of $\epsilon^{SR}$, one sees that this object is proportional to the inverse of the derivative of the potential:
	\begin{equation}
		\mathcal{P}\propto \frac{1}{(U'(\chi))^2} \,.
		\label{PowSR2}
	\end{equation}
	Then, one can wonder if a flat region of the potential can lead to a large enhancement of the curvature perturbation \cite{Garcia-Bellido:2017mdw}.
	
	As it is pointed out in \cite{Germani:2017bcs}, a very flat region in the potential would lead the inflationary dynamics to leave the slow roll attractor for the $ultra$ $slow$ $roll$ (USR) regime \cite{Tsamis:2003px,Kinney:2005vj}. Following, e.g.~\cite{Germani:2017bcs}, we shall present the main features of the USR dynamics by mean of simple analytic arguments. 
	The pure USR regime can be defined by a perfectly flat potential. The background Klein-Gordon equation can then be written as
	\begin{equation}
		\chi'' + \left(3-\epsilon_H\right)\chi' = 0 \,,
		\label{EOMUSR}
	\end{equation}
	where $\epsilon_H$ is the geometric SR parameter. It follows from its definition in \eq \eqref{defHdefEps} that
	\begin{equation}
		\epsilon_H(N) = -\frac{H'(N)}{H(N)} \,.
		\label{DefEpsGeom}
	\end{equation}
	On the other hand, the first of the Friedman equations \eqref{Friedman} can be written as:
	\begin{equation}
		\iff H^2 = \frac{2 U}{6-(\chi')^2} \,.
		\label{HUSR}
	\end{equation}
	If one assumes that the USR was preceded by a perfect SR evolution, one can initially neglect the parameter $\epsilon_H$. Thus, equation \eqref{EOMUSR} can be integrated and the inflaton's velocity is exponentially suppressed in time:
	\begin{equation}
		\chi' \sim e^{-3 N} \;.
		\label{ChisolUSR}
	\end{equation}
	By combining the definition \eqref{DefEpsGeom} and the dynamical equation \eqref{HUSR}, one can express $\epsilon_H$ as function of the field's velocity\footnote{Equation \eqref{epsBeyond} is completely general, it also applies in other regimes.}:
	\begin{equation}
		\epsilon_H = \frac{(\chi')^2}{2} \;.
		\label{epsBeyond}
	\end{equation}
	Using the solution \eqref{ChisolUSR}, this can be used to express $\epsilon_H$:
	\begin{equation}
		\epsilon_H\sim e^{-6 N} \;.
		\label{EpsUSR}
	\end{equation}
	The first slow roll parameter is then even smaller in this situation than during the SR evolution. 
	
	However, this is completely different for the second slow-roll parameter $\eta_H$, which according to \eq \eqref{etaEpsilon} reads
	\begin{equation}
		\eta_H(N) = \epsilon_H(N)-\frac{1}{2}\frac{\epsilon_H'(N)}{\epsilon_H(N)}\;.
	\end{equation}
	By making use of \eq \eqref{EpsUSR}, one can compute $\eta_H$ in the leading order of USR approximation and
	\begin{equation}
		\eta_H = 3 \;.
	\end{equation}
	It is a clear violation of the second slow roll condition $\eta_H\ll 1$. This could be understood qualitatively. Indeed, when the contribution of the potential's slope becomes negligible, the main contribution to the motion will come from the Hubble friction term. Thus, because of the strong falloff of the field's velocity, it becomes wrong to neglect its acceleration.
	All of this provides us another argument pointing toward a large enhancement of the power spectrum. By plugging equation \eqref{EpsUSR} into the curvature power spectrum \eqref{PowSR}, one sees that it is indeed strongly enhanced: 
	\begin{equation}
		\mathcal{P}_\mathcal{R}\sim e^{6 N} \;.
		\label{EnhanceUSR}
	\end{equation}
	We see that this allows a huge -- and non-divergent -- enhancement of $\mathcal{P}$ for exactly flat regions of the potential. Moreover, we see that the duration of USR phase will determine the strength of this enhancement.

\section{Palatini Higgs Inflation} \label{sec:HiggsInflation}

\subsection{The theory}
 Let us now specify our discussion to the particular case of Higgs inflation. In both the metric and Palatini formulations, the action of Higgs inflation is given by \cite{Bezrukov:2007ep, Bauer:2008zj}
\begin{equation}
	S = \frac{1}{2}\int \diff^4 x\sqrt{-g}\left\{\Omega^2(h)R - g^{\mu\nu}\partial_\mu h\partial_\nu h - 2 V(h)\right\} \;.
	\label{Act1}
\end{equation} 
Here, $R$ represents the Ricci scalar, and we have employed the unitary gauge for the Higgs field. Furthermore,
\begin{equation}
	\Omega^2(h) = 1+\xi h^2\;, \qquad V(h) = \frac{\lambda}{4}(v^2-h^2)^2\sim \frac{\lambda}{4}h^4\;,
\end{equation}
where $\xi$ denotes the strength of the non-minimal coupling of $h$ to gravity, and we have neglected the vacuum expectation value $v$ in the Higgs potential.
To analyse the action \eqref{Act1}, we have to put it in a form similar to \eqref{ActBackField}. To this aim, we perform a Weyl transformation to shift to the Einstein frame \cite{Bezrukov:2007ep,Bauer:2008zj}:
\begin{equation}
	g_{\mu\nu} = \Omega^{-2}(h)\Tilde{g}_{\mu\nu}\;, \qquad \sqrt{-g} = \Omega^{-4}(h)\sqrt{-\Tilde{g}}\;, \qquad g^{\mu\nu} = \Omega^2(h)\Tilde{g}^{\mu\nu}\;.
	\label{conformal}
\end{equation}
Consequently, we obtain
\begin{equation}
	S = -\frac{1}{2}\int \diff^4 x\sqrt{-\Tilde{g}}\left\{\Tilde{R}+K(h)\Tilde{g}^{\mu\nu}\partial_\mu h\partial_\nu h + 2\Omega^{-4}(h)V(h)\right\} \;,
\end{equation}
where the explicit form of $K(h)$ now depends on the formulation of GR. In this work, we will mainly focus on the Palatini version, in which the metric $g_{\mu\nu}$ and the connection $\Gamma^\alpha_{\beta \gamma}$ are considered independent. In this scenario, the matter sector will typically source torsion $T^\alpha_{\beta\gamma} = \Gamma^\alpha_{\beta \gamma} - \Gamma^\alpha_{\gamma\beta}$ and non-metricity $Q_{\alpha\beta\gamma} = \nabla_\alpha g_{\beta \gamma}$. We can consider only $T^\alpha_{\beta\gamma}$, only $Q_{\alpha\beta\gamma}$, or both $T^\alpha_{\beta\gamma}$ and $Q_{\alpha\beta\gamma}$ simultaneously. For the action \eqref{Act1}, all three options are equivalent \cite{Rigouzzo:2022yan}.

Since an independent connection renders the Ricci tensor conformally invariant, the transformation \eqref{conformal} is straightforward, and the kinetic function $K$ is then given by
\begin{equation}
	K = \frac{1}{\Omega^2(h)} \;.
	\label{KPal}
\end{equation}
Next, it is convenient to canonically normalize the kinetic term by defining the field $\chi$ such that:
\begin{equation}
	\sqrt{K(h)} \diff h = \diff \chi  \qquad \implies \qquad \chi(h) = \int^h_0 \diff x\sqrt{K(x)} \;.
	\label{DefCan}
\end{equation}
The final action is then given by:
\begin{equation}
	S = -\frac{1}{2}\int \diff^4 x\sqrt{-\Tilde{g}}\left\{\Tilde{R} + \Tilde{g}^{\mu\nu}\partial_\mu\chi\partial_\nu\chi + 2 U(\chi)\right\}\;,
	\label{EinsteinFrame}
\end{equation}
where the Einstein frame potential $U(\chi)$ is expressed as
\begin{equation}
	U(\chi) = \frac{V(h(\chi))}{\Omega^4(h(\chi))} = \frac{\lambda}{4}\left(\frac{h(\chi)}{\Omega(h(\chi))}\right)^4\;.
\end{equation}
To simplify the notation, we can define the function $F(\chi)$ as:
\begin{equation}
	F(\chi) = \frac{h(\chi)}{\Omega(h(\chi))} \;,
	\label{DefF}
\end{equation}
which leads to the Einstein frame potential
\begin{equation}
	U(\chi) = \frac{\lambda}{4}F^4(\chi) \;.
	\label{PotF}
\end{equation}
In the case of the Palatini formulation, $F$ takes a simple analytic form:
\begin{equation}
	F(\chi) = \frac{1}{\sqrt{\xi}}\tanh(\sqrt{\xi}\chi) \;.
	\label{FPalatini}
\end{equation}
The field being minimally coupled with a well-defined potential, one can carry on the standard analysis described in section \ref{sec:GenInfl}.

\subsection{Tree-Level Inflation}

First, we shall review Palatini Higgs inflation to the leading order in the slow-roll approximation. Our computation will otherwise be exact; in particular, we shall not assume that $\xi$ is large. The first slow-roll parameter is then given by
\begin{equation}
	\epsilon^{\text{SR}} = \frac{1}{2}\left(\frac{U'(\chi)}{U(\chi)}\right)^2 =  \frac{32 \xi}{
		\sinh^2\left(2\sqrt{\xi} \chi\right)} \;. 
	\label{EpsPalTree}
\end{equation}
As discussed in section \ref{sec:GenInfl}, we define the end of inflation by a violation of the first SR condition \eqref{SR1}\cite{Gorbunov:2011zzc, Riotto:2018pcx}
\begin{equation}
	\epsilon^{SR}(\chi_{\text{end}}^{\text{SR}})= 1 \;,
\end{equation}
which can be solved using \eqref{EpsPalTree}:
\begin{equation}  \label{ChiEndSR}
	\chi_{\text{end}}^{\text{SR}} = \frac{\arcsinh\left(\sqrt{32\xi}\right)}{2\sqrt{\xi}} \;.
\end{equation}
Now, one can use the slow-roll approximation to solve the equations of motion for the number of e-foldings $N$:
\begin{equation} \label{eqN}
	N = \int^{\chi^{\text{SR}}_{\text{in}}}_{\chi^{\text{SR}}_N} \mathrm{d}\chi \frac{U(\chi)}{U'(\chi)} = \int^{\chi^{\text{SR}}_{\text{in}}}_{\chi^{\text{SR}}_N} \mathrm{d}\chi \frac{\sinh\left(2\sqrt{\xi}\chi\right)}{8\sqrt{\xi}} = \frac{1}{16\xi}\left(\cosh\left(2\sqrt{\xi}\chi^{\text{SR}}_{\text{in}}\right)-\cosh\left(2\sqrt{\xi}\chi^{\text{SR}}_N\right)\right) \;,
\end{equation}
where $\chi^{\text{SR}}_{\text{in}}$ and $\chi^{\text{SR}}_N$ correspond to the field value at some initial moment and $N$ e-foldings later, respectively.

Defining $N_\star$ as the number of e-foldings between the generation of perturbations observed in the CMB and the end of inflation, we have $\chi^{\text{SR}}_{N_\star} = \chi^{\text{SR}}_{\text{end}}$. Inverting \eqref{eqN} yields
\begin{equation} \label{chiCMB}
	\chi^{\text{SR}}_{\text{CMB}} = \frac{\arccosh\Big(16 \xi N_\star+\cosh\left(2\sqrt{\xi}\chi^{\text{SR}}_{\text{end}}\right)\Big)}{2 \sqrt{\xi}} = \frac{\arccosh\Big(16 \xi N_\star+\sqrt{32\xi + 1}\Big)}{2 \sqrt{\xi}} \;.
\end{equation}
Now we can evaluate relevant quantities at the time corresponding to the CMB:
\begin{equation} \label{FCMB}
	F^{\text{SR}}_{\text{CMB}} = \frac{1}{\sqrt{\xi}} \tanh\left(\frac{1}{2} \arccosh\left(16 \xi  N_\star+\sqrt{32 \xi +1}\right)\right) \;,
\end{equation}
which gives
\begin{equation} \label{epsilonCMB}
	\epsilon^{\text{SR}}_{\text{CMB}} = \frac{1}{N_\star \left(\sqrt{32 \xi +1}+8 \xi  N_\star\right)+1} \;,
\end{equation}
corresponding to 
\begin{equation} \label{potentialCMB}
	U^{\text{SR}}_{\text{CMB}} =  \frac{\lambda}{4 \xi^2} \tanh^4\left(\frac{1}{2} \arccosh\left(16 \xi  N_\star+\sqrt{32 \xi +1}\right)\right) \;.
\end{equation}
Finally, we can compute the amplitude of CMB perturbations:
\begin{equation} \label{normalizationCMB}
	\frac{U^{\text{SR}}_{\text{CMB}}}{\epsilon^{\text{SR}}_{\text{CMB}}} = \frac{\lambda}{4 \xi^2} \left(N_\star \left(\sqrt{32 \xi +1}+8 \xi  N_\star\right)+1\right) \tanh^4\left(\frac{1}{2} \arccosh\left(16 \xi  N_\star+\sqrt{32 \xi +1}\right)\right) \;,
\end{equation}
which should be compared with the observed value \eqref{PLANCK} in the CMB \cite{Akrami:2018odb}.

We can solve \eqref{normalizationCMB} for the non-minimal coupling $\xi$. Since the result will show that $\xi \gg 1$, we obtain
\begin{equation} \label{normalizationCMBTree}
	\frac{U^{\text{SR}}_{\text{CMB}}}{\epsilon^{\text{SR}}_{\text{CMB}}} \approx \frac{2\lambda N_\star^2}{\xi} = 5 \times 10^{-7} \quad \implies \quad \xi = \frac{2}{5}\lambda N_\star^2 \times 10^7 = 1.21 \times 10^{10}\lambda \;,
\end{equation}
where we used $N_\star = 55$ as a typical value (see \cite{Shaposhnikov:2020fdv}).\footnote{A more precise value of $N_\star$ can be deduced from knowledge of preheating \cite{Rubio:2019ypq, Dux:2022kuk}.}
This shows that if the tree-level analysis is valid, then the non-minimal coupling $\xi$ has to be large, $\xi \sim 10^7$.

\subsection{RG Running within the Standard Model}
However, it is well-known that the predictions of Palatini Higgs inflation can be crucially altered by quantum effects \cite{Bauer:2010jg,Rasanen:2017ivk, Markkanen:2017tun,Enckell:2018kkc,Jinno:2019und,Shaposhnikov:2020fdv,Enckell:2020lvn}. One of them is the renormalization group (RG) running of the coupling constants. Based on investigations in the metric scenario \cite{Bezrukov:2007ep,DeSimone:2008ei,Bezrukov:2008ej,Bezrukov:2009db,Barvinsky:2009fy,Barvinsky:2009ii,Hamada:2014iga,Bezrukov:2014bra,Bezrukov:2014ipa,Fumagalli:2016lls,Enckell:2016xse,Bezrukov:2017dyv}, this effect has already been studied in the Palatini case \cite{Rasanen:2017ivk,Markkanen:2017tun,Enckell:2018kkc,Shaposhnikov:2020fdv,Enckell:2020lvn}. The running Higgs self-coupling $\lambda(\mu)$ can be fitted with a well-known function \cite{Isidori:2007vm}
\begin{equation}
	\lambda(\mu) = \lambda_0 + b \log^2\left(\frac{\mu}{q}\right) \;,
	\label{LogFit}
\end{equation}
which we shall employ for throughout our study.
Here, $\mu$ is the renormalization scale, and the constants $\lambda_0$, $b$, and $q$ are determined by the parameters of the Standard Model (SM) as measured at low energies. The most important contribution to the RG running of $\lambda$ comes from the interaction with the top quark 
\begin{equation} \label{yukawaJordan}
	\mathcal{L}_{\text{top}} = \frac{y_t}{\sqrt{2}}\, h\,\Bar{\psi}_t\psi_t \;,
\end{equation}
where $y_t$ is the top Yukawa coupling. However, a sizable uncertainty in $\lambda(\mu)$ arises due to the finite precision of measurements of its low-energy value, $y_t^{\text{low}}$ (see detailed discussion in \cite{Bezrukov:2014ina}). In the $\overline{\text{MS}}$-scheme and at the normalization point $\mu = 173.2\,\text{GeV}$. The parameters of the fit are usually around the following typical values \eqref{LogFit}\footnote
{We thank Fedor Bezrukov for kindly providing us with Mathematica notebooks, which match physical and $\overline{\text{MS}}$-parameters at two loops \cite{Bednyakov:2015sca} and compute the three-loop RG running of the coupling constants \cite{Chetyrkin:2012rz} (see also \cite{Bezrukov:2012sa,Degrassi:2012ry,Buttazzo:2013uya}).
As an input, we use $\alpha_s(M_Z)=0.1181$ and $m_h=125.1\,$GeV. We will not consider the effect of $\xi$ since its influence on the running is negligible \cite{Bezrukov:2009db}.}
\begin{equation}
	q\sim \mathcal{O}(0.3)\;,\qquad b\sim \mathcal{O}(2\times 10^{-5})\;,\qquad |\lambda_0|\ll 1\;.
	\label{ParamSpace}
\end{equation}
As one can immediately see, the minimal value of the quartic coupling, $\lambda_0$, can vary significantly and can assume both positive and negative values. 

Finally, we need to determine $\mu$ at inflationary field values. Since the top quark has the strongest influence on the running of $\lambda$, we can employ the principle of minimal sensitivity \cite{Stevenson:1981vj,Stevenson:1982wn} to identify $\mu$ with the high-energy value of the top quark mass \cite{Bezrukov:2014ipa,Shaposhnikov:2020fdv}. Transforming \eq \eqref{yukawaJordan} to the Einstein frame, we get 
\begin{equation}
	\mathcal{L}_{\text{top}} =  \frac{y_t}{\sqrt{2}}\, F(\chi)\Bar{\psi}_t\psi_t \;.
\end{equation}
This leads to the field-dependent renormalization scale
\begin{equation}
	\mu(\chi)= y_t F(\chi).
	\label{EnergyScale}
\end{equation}
For the high-energy value of the top Yukawa, we will follow \cite{Shaposhnikov:2020fdv} and approximate $y_t\sim 0.43$. In the following, $y_t$ without superscript shall always refer to high energies. As a result, the Higgs potential including the effects of RG running reads
\begin{equation}
	U(\chi) = \frac{\lambda(\mu)}{4}F^4(\chi) = \frac{1}{4} \left(\lambda_0 + b \log^2\left(\frac{y_t F(\chi)}{q}\right) \right) F^4(\chi) \;.
\end{equation}
\subsection{Additional Contributions from UV-Physics}
So far, we have computed the renormalization group (RG) running assuming that there are no contributions beyond the Standard Model (SM). We will now relax this assumption and take into account possible effects of unknown physics at high energies. In the metric formulation of General Relativity (GR), it has been shown that renormalizing the effective potential at one loop leads to \cite{Bezrukov:2014ipa}:\footnote
{To avoid any confusion, please note that our $\delta\lambda$ has the opposite sign compared to the one in \cite{Bezrukov:2014ipa}. We made this choice for later convenience. We also note that in general one can include a similar contribution for the top Yukawa coupling \cite{Bezrukov:2014ipa}, which we will not consider.}
\begin{equation} \label{shiftLambda}
	\lambda(\mu) \to \lambda_{SM}(\mu) + \delta\lambda\left(1-G^2(\chi)\right) \;,
\end{equation}
where $\delta \lambda$ is an unknown parameter and 
\begin{equation}
	G(\chi) = (F'(\chi))^2 + \frac{F''(\chi)F(\chi)}{3} \;.
	\label{DefG}
\end{equation}
For $\chi=0$, $G$ is of order one, whereas $\chi\rightarrow \infty$ leads to a vanishing $G$. Therefore, $\delta \lambda$ represents a jump of the coupling constant between the low- and high-energy regimes.

Subsequently, this study was extended to Palatini Higgs inflation \cite{Shaposhnikov:2020fdv}: The form of possible corrections as shown in \eqs \eqref{shiftLambda} and \eqref{DefG} remains the same, but the effects of unknown UV-physics are expected to be less important compared to the metric formulation. A rough estimate shows $\delta \lambda \lesssim 10^{-3}$, where the smallness of this numerical value is mainly due to loop suppression \cite{Shaposhnikov:2020fdv}. Since $\lambda(\mu)$ as computed within the SM can approach very small values at high energies, corresponding to $|\lambda_0| \ll 1$, an effective jump $\delta \lambda$ on the order of $10^{-3}$ can nevertheless have a significant effect, as we shall discuss. An analogous argument can be applied to estimate the numerical value of a possible jump $\delta y_t$ of the top Yukawa coupling. Since $y_t$ stays on the order of $1$ also at high energies, we expect the effect of such a small $y_t$ to be negligible and we will not consider $\delta y_t$ it in the following.

Plugging in the specific form \eqref{FPalatini} of $F$ in Palatini HI, we get 
\begin{equation} \label{GConcrete}
	G(\chi) = \frac{5-2\cosh^2(\sqrt{\xi}\chi)}{3\cosh^4(\sqrt{\xi}\chi)} \;.
\end{equation}
Thus, the potential reads
\begin{equation}
	U(\chi) = \frac{1}{4} \left(\lambda_0 + b \log^2\left(\frac{y_t F(\chi)}{q}\right) + \delta\lambda\Big(1-G^2(\chi)\Big)  \right)F^4(\chi)\;.
	\label{FullPot}
\end{equation}
Its derivative is given by 
\begin{equation}
	\begin{split}
	U'(\chi) = \left(\lambda_0 + b \log^2\left(\frac{y_t F(\chi)}{q}\right) +\frac{b}{2}\log\left(\frac{y_t F(\chi)}{q}\right)\right)F'(\chi)F^3(\chi)\\ + \delta\lambda\left((1-G^2(\chi))\frac{F'(\chi)}{F(\chi)} - \frac{G'(\chi)G(\chi)}{2}\right)F^4 \;.
	\end{split}
	\label{Derivative}
\end{equation}
We note that a potential in the form of \eqs \eqref{FullPot} and \eqref{Derivative} does not include an influence of $\delta \lambda$ on the parameters $q$, $b$ and $\lambda_0$, which is a good approximation for small $\delta \lambda$.

At this point, we can discuss the relationship of our approach to the previous studies \cite{Rasanen:2018fom,Enckell:2020lvn}, where critical points in Palatini HI were found.\footnote
{We thank Syksy R\"as\"anen and Eemeli Tomberg for very useful correspondence.}
An important difference is that an effective jump $\delta y_t$ of the top Yukawa coupling was included in \cite{Rasanen:2018fom,Enckell:2020lvn} and indeed values $|\delta y_t| \gtrsim 10^{-2}$ led to features in the inflationary potential. Moreover, the jumps in \cite{Rasanen:2018fom} occur at lower field values $\chi \sim M_P/\xi$ whereas in our framework they are located around $\chi \sim M_P/\sqrt{\xi}$. The influence of jumps can increase if they happen at lower field values since then RG running is altered over larger field ranges. As another aspect, a jump at $\chi \sim M_P/\xi$ is effectively degenerate with a change of the low-energy values of coupling constants \cite{Rasanen:2018fom,Enckell:2020lvn}. This is no longer the case in our approach: a change of the low-energy value of $\lambda$ generically has a different effect than a jump $\delta \lambda$. In other words, we see effects of the jump happening during inflation. Finally, instantaneous jumps were employed in \cite{Rasanen:2018fom,Enckell:2020lvn} instead of the interpolation \eqref{DefG}.

\section{Parameter Space with a Large Enhancement of Perturbations}
\label{ParamSpa}

As stated above, our goal is to determine whether large perturbations can be generated from a phase of ultra slow-roll due to the presence of an inflection point or a shallow local minimum in the Einstein frame potential. Thus, we will look for regions of the $(\xi,\lambda_0,\delta\lambda)$ space such that the potential retains this particular shape, maintaining its positivity and asymptotic flatness. To find the stationary points of the potential \eqref{FullPot}, we need to find the zeros of its derivative \eqref{Derivative}, which leads to the following master equation:
\begin{equation}
	\left(\Bar{\lambda}_0 + \frac{b}{2}\log\left(\frac{\mu(\chi)}{q}\right) + b \log^2\left(\frac{\mu(\chi)}{q}\right)\right) = \delta\lambda \frac{F(\chi)}{F'(\chi)}\left(\frac{G'(\chi)G(\chi)}{2} + G^2(\chi) \frac{F'(\chi)}{F(\chi)}\right)\;,
	\label{MasterEq}
\end{equation}
where we have defined the shifted coupling by
\begin{equation}
	\Bar{\lambda}_0=\lambda_0 + \delta\lambda\;.
\end{equation}
In what follows, we will determine which regions of the parameter space are permissible in order to solve \eqref{MasterEq} and will demonstrate that it cannot lead to a consistent inflationary scenario.

\subsection{Jump-less Scenario}
\label{ssec:jumpless}
First, let us consider the case where there is no jump in the coupling constant, i.e., the limit where $\delta\lambda = 0$. We are then left with a two-dimensional parameter space which will be simpler to study. Moreover, it will become apparent below that the jumping case is not so different in most of the parameter space.

Equation \eqref{MasterEq} simply becomes:
\begin{equation}
	\lambda_0 + \frac{b}{2}\log\left(\frac{\mu}{q}\right) + b \log^2\left(\frac{\mu}{q}\right)=0\;.
	\label{JumpLess}
\end{equation}
This can be solved explicitly for $\mu$, with the two solutions given by:
\begin{equation}
	\mu_\pm = q E_\pm,\quad E_\pm = \exp\left\{\frac{1}{4}\left(-1\pm\sqrt{1-\frac{16\lambda_0}{b}}\right)\right\}\;.
	\label{Sol1}
\end{equation}
To obtain two real solutions, the following constraint on $\lambda_0$ must be imposed:
\begin{equation}
	\lambda_0 < \frac{b}{16}\sim 1.25\times 10^{-6}\;.
	\label{Const1}
\end{equation}
Let us stress that we have not made any assumptions about the formulation of gravity here. Indeed, the specificity of each formulation is actually contained in the particular shape of the function $F$. The bound \eqref{Const1} is then a universal property of the Higgs inflation effective potential.

The next step to finding the location of the stationary points is to invert the solution \eqref{Sol1} using the definition \eqref{EnergyScale} of the inflationary scale $\mu(\chi)$:
\begin{equation}
	\chi_\pm = F^{-1}\left(\frac{q E_\pm}{y_t}\right)\;.
	\label{FormalChiPm}
\end{equation}
The advantage of the Palatini formulation is the simple analytic form of the function $F$ given by \eqref{FPalatini}. It is then possible to provide an explicit expression for $\chi_\pm$, which may not be possible in more complicated gravity formulations. The non-zero extrema of $U(\chi)$ are then given by:
\begin{equation}
	\chi_\pm = \frac{1}{\sqrt{\xi}}\text{arctanh}\left(\frac{\sqrt{\xi}q E_\pm}{y_t}\right)\;.
	\label{PalatniChiPm}
\end{equation}
From \eqref{PalatniChiPm}, one can automatically derive a bound for the non-minimal coupling by requiring the $\tanh$ argument to be smaller than one. Thus:
\begin{equation}
	\chi_\pm \in \mathbb{R}\implies \xi < \xi_{max} = \left(\frac{y_t}{q E_+}\right)^2\;.
	\label{Const2Pal}
\end{equation}

To derive the bound \eqref{Const2Pal}, we used the precise form of $F$ in the Palatini formulation \eqref{FPalatini}. However, since \eqref{Const1} was actually formulation-independent, one may ask if the same thing can be said about \eqref{Const2Pal}. The answer to this question is yes: this constraint is actually independent of the gravity formulation one chooses.

To illustrate this, note that it is possible to find a formal expression for the function $F^{-1}$ without assuming a particular form of the kinetic function $K(h)$. Indeed, if one applies $F^{-1}$ to both sides of equation \eqref{DefF} and uses the definition \eqref{DefCan}, we are left with:
\begin{equation}
	\chi(h)= F^{-1}\left(\frac{h}{\sqrt{1+\xi h^2}}\right) = \int^h_0 d x \sqrt{K(x)}\;.
\end{equation}
Then, by redefining the variable $h$, one can write:
\begin{equation}
	F^{-1}(y) = \int^{\frac{y}{\Omega(i y)}}_0 d x\sqrt{K(x)}\;.
	\label{IntegrF}
\end{equation}
To ensure the realness of the upper bound of the integral \eqref{IntegrF}, one has to impose that $y < 1/\sqrt{\xi}$. This bound corresponds to the domain where the function $F$ is invertible. This is similar to the situation in Palatini, where we imposed that $\sqrt{\xi}q E_+/y_t$ must be within the definition domain of the arctanh. Thus, to maintain the realness of the potential's stationary point, one must ensure that the argument of $F^{-1}$ in \eqref{FormalChiPm} satisfies this constraint. This yields the same upper bound as in \eqref{Const2Pal}:
\begin{equation}
	\Omega(i\frac{q E_\pm}{y_t})\in\mathbb{R}\implies \xi < \xi_{max} = \left(\frac{y_t}{q E_+}\right)^2\;.
	\label{Const2}
\end{equation}
To estimate this upper bound, note from \eqref{Sol1} that $E_+\geq e^{-1/4}$. Then, using \eqref{ParamSpace}, one can see that
\begin{equation} \label{xiMaxInflection}
	\xi_{max}\sim \left(\frac{1}{E_+}\right)^2 \lesssim \left(\frac{1}{e^{-1/4}}\right)^2\sim \mathcal{O}(1)\;.
\end{equation}

In summary, for any formulation of general relativity, the allowed parameter space to observe a local minimum in the inflaton's potential in the jumpless situation is $\xi\lesssim \mathcal{O}(1)$ and $\lambda_0\lesssim\mathcal{O}(10^{-6})$. As we will show below, these constraints are not compatible with the PLANCK normalization of the power spectrum.
\subsection{Scenario with the Jump}
\label{ssec:withJump}
In this section, we take into account the contributions from UV physics. As mentioned above, this manifests itself in practice by adding an arbitrary jump, $\delta\lambda$, to the running of the coupling constant.\\
\begin{minipage}{0.45\linewidth}
	\begin{figure}[H]
		\hspace{-0.9cm}
		\includegraphics[scale = 0.6]{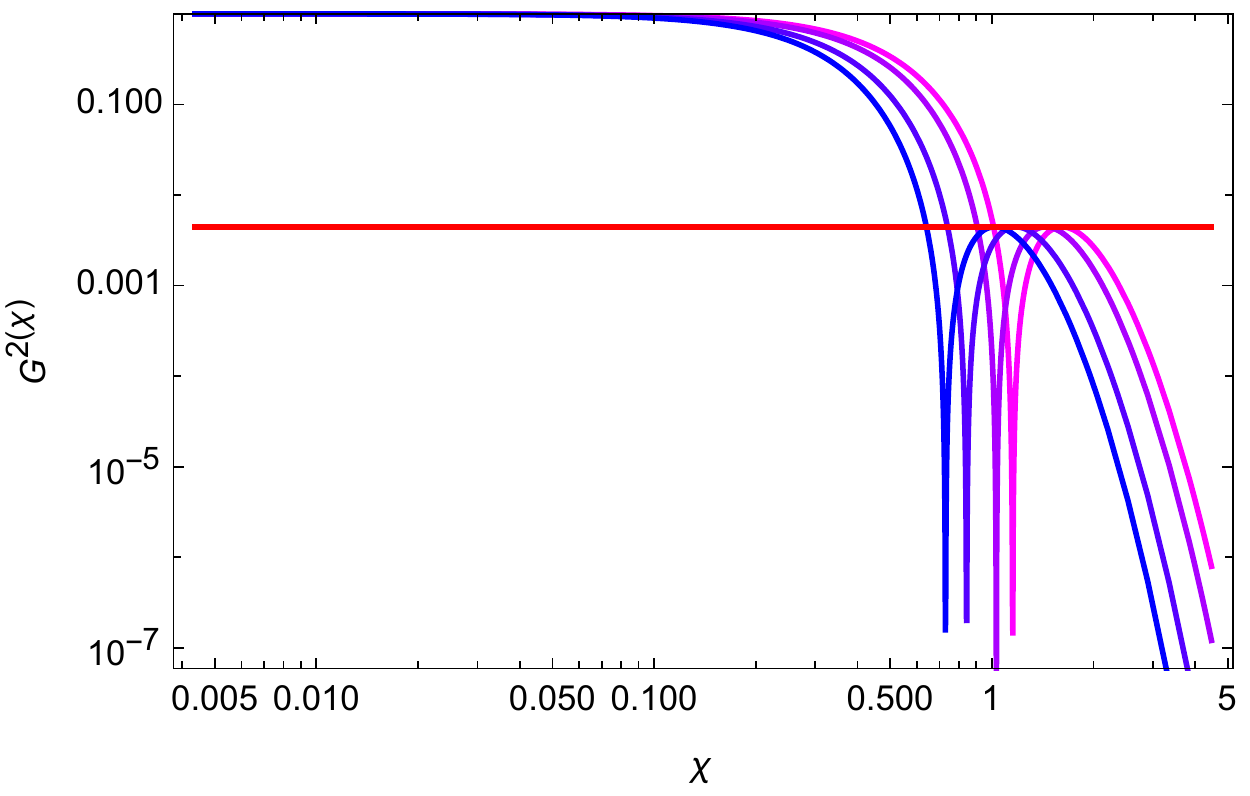}
		\caption{Function $G^2(\chi)$ as function of the field for different values of $\xi$: (from right to left) $\xi=0.8$, $1$, $1.5$, $2$.}
		\label{Gsquare}
	\end{figure}
\end{minipage}
\hfill
\begin{minipage}{0.45\linewidth}
	Our parameter space is then, a priori, unconstrained and three-dimensional. We will begin by analytically studying the behaviour of the function $G^2$, which is defined in equation \eqref{GConcrete}. As shown in Figure \ref{Gsquare}, $G^2$ has a global maximum at $\chi = 0$, as $G^2(\chi=0) = 1$, and two stationary points at
	\begin{equation}
		\begin{split}
			\chi_1 = \frac{1}{\sqrt{\xi}}\text{arccosh}\left(\sqrt{\frac{5}{2}}\right)\sim \frac{1.03}{\sqrt{\xi}},\\
			\chi_2 = \frac{1}{\sqrt{\xi}}\text{arccosh}\left(\sqrt{5}\right)\sim \frac{1.4}{\sqrt{\xi}}\;,
		\end{split}
	\end{equation}
	with $\xi$ independent values:
\end{minipage}

\begin{equation}
	G^2(\chi_1) = 0\text{ and }G^2(\chi_2) = 1/225\;.
	\label{LittleBump}
\end{equation}
For $\chi>\chi_2$, the function is exponentially suppressed as:
\begin{equation}
	\lim_{\chi\gg\chi_2}G^2(\chi) = \frac{64}{9}e^{-4\sqrt{\xi}\chi} \to 0\;.
\end{equation}
Consequently, we will neglect the effect of the function $G^2$ for field values $\chi > \chi_2$.
Now, let us interpret these results by comparing them with the endpoint of inflation in the tree-level slow-roll approximation \eqref{ChiEndSR}.
The condition for the jump to occur before the end of inflation yields:
\begin{equation} \label{xiMaxJump}
	\chi_{\text{end}}^{\text{SR}} < \chi_1 \iff \xi < \xi_{\text{thresh}} =\frac{1}{32} \sinh ^2\left(2 \arccosh\left(\sqrt{\frac{5}{2}}\right)\right)\sim 0.47 \;,
\end{equation}
which excludes most of the parameter space.

Now, we need to separately study the cases where the jump occurs during inflation and the contrasting situation where the coupling constant is only shifted by a constant. To avoid any confusion, let's recall that the bounds \eqref{Const1} and \eqref{Const2} do not apply \emph{a priori} to the case involving a jump, i.e., we have placed no constraints on the $(\xi,\lambda_0)$ space so far.
\subsubsection{If The Jump Does Not Occur During Inflation}
For $\xi > \xi_{thresh}$, one can safely neglect the effect of the function $G^2$ in equation \eqref{MasterEq}, which reduces to:
\begin{equation}
	\Bar{\lambda}_0 + \frac{b}{2}\log\left(\frac{\mu}{q}\right) + b \log^2\left(\frac{\mu}{q}\right) = 0\;.
	\label{LocMinJump}
\end{equation}
Remarkably, one can notice that the situation is exactly similar to the jumpless case \eqref{JumpLess}, except for the replacement $\lambda_0\to\Bar{\lambda}_0$. Thus, one can repeat the analysis of the previous section to generalize the constraints \eqref{Const1} and \eqref{Const2}:
\begin{equation}
	\Bar{\lambda}_0 < \frac{b}{16}\sim 1.25\times 10^{-6},\quad \xi_{thresh}\lesssim\xi<\xi_{max} = \left(\frac{y_t}{q E_\pm}\right)^2\sim \mathcal{O}(1)\;.
	\label{Const3}
\end{equation}
Note that this is only possible in the case where $\xi_{thresh}>\xi_{max}$, which yields additional constraints on the jump parameter. However, it will be sufficient for our purpose to use the bound \eqref{Const3} on $\Bar{\lambda_0}$ and the order of magnitude of the non-minimal coupling.
\subsubsection{If The Jump Does Occur During Inflation}
For smaller $\xi$, one cannot neglect the right-hand side (RHS) of equation \eqref{MasterEq} for small values of the field. The analytic study of the system in this case may be a bit trickier. However, it is still possible to get some understanding of what happens due to the simple form of $F$ in the Palatini formulation. To be more transparent, let us package the RHS contributions of \eqref{MasterEq} in a function $A(\chi)$ defined as:
\begin{equation}
	A(\chi) =\frac{F(\chi)}{F'(\chi)}\left(\frac{G'(\chi)G(\chi)}{2} + G^2(\chi) \frac{F'(\chi)}{F(\chi)}\right)
\end{equation}
\begin{equation}
	=
	\frac{\left(5-2\cosh^2(\sqrt{\xi}\chi)\right)\left(15 - 14 \cosh^2(\sqrt{\xi}\chi) + 2\cosh^4(\sqrt{\xi}\chi)\right)}{9\cosh^{8}(\sqrt{\xi}\chi)}\;.
\end{equation}
This function can be studied precisely. It has $4$ extrema at
\begin{equation}
	\chi = 0,\quad0.702/\sqrt{\xi},\quad1.219/\sqrt{\xi}\text{ and }1.996/\sqrt{\xi}\;.
\end{equation}
Because of the $1/\sqrt{\xi}$ form of these roots, the extreme values of $A$ are universal for all $\xi$. Thus, one can see that $\chi = 0$ is the global maximum of the function with $A(\chi=0) = 1$. Therefore, we can state that
\begin{equation}
	A(\chi) < 1,\quad \forall\chi >0.
\end{equation}
Now, we can rewrite the master equation \eqref{MasterEq} as:
\begin{equation}
	\lambda_0 + \frac{b}{2}\log\left(\frac{\mu(\chi_\star)}{q}\right) + b \log^2\left(\frac{\mu(\chi_\star)}{q}\right)= \delta\lambda(A(\chi)-1)\;.
	\label{MasterEqH}
\end{equation}
\begin{minipage}{0.45\linewidth}
	\begin{figure}[H]
		\hspace{-0.9cm}
		\includegraphics[scale = 0.6]{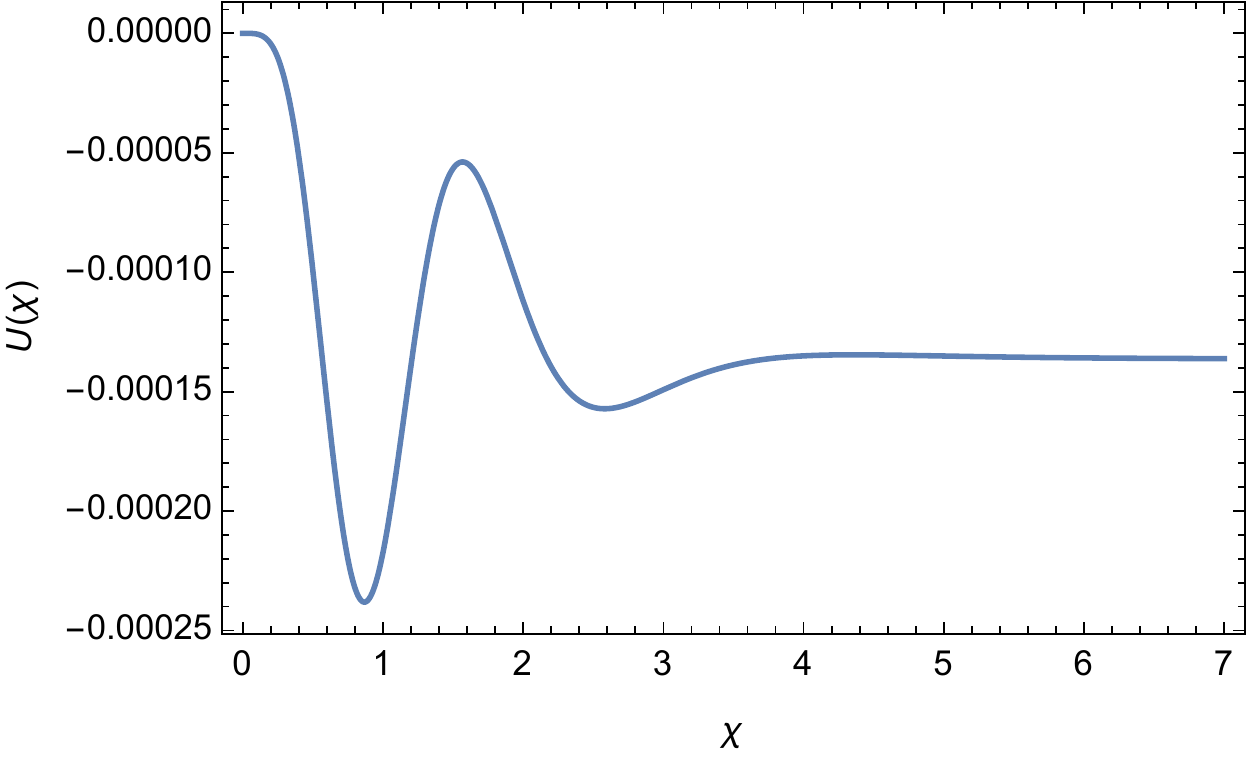}
		\caption{Potential \eqref{FullPot} with the tuned jump using \eqref{deltaLambdaPos} for $\xi = 0.4$ and $y_t^{low} = 0.934$. One can see that the potential has a local minimum but goes through negative values, which makes it unstable and improper for inflation.}
		\label{Wrong}
	\end{figure}
\end{minipage}
\hfill
\begin{minipage}{0.5\linewidth}
	If a solution $\chi_\star$ to \eqref{MasterEqH} exists, then the corresponding jump parameter will take the form:
	\begin{equation}
		\delta\lambda_{|\chi_\star} = \frac{\lambda_0 + \frac{b}{2}\log\left(\frac{\mu(\chi_\star)}{q}\right) + b \log^2\left(\frac{\mu(\chi_\star)}{q}\right)}{A(\chi_\star)-1}\;.
		\label{deltaLambdaPos}
	\end{equation}
	However, this is not sufficient for consistency with Higgs inflation. Indeed, this prescription only provides us a local minimum in the potential before the slow-roll (SR) end of inflation, but it does not tell us anything about the sign of the potential or its asymptotic behaviour. One can perfectly tune the $\delta\lambda_\star$ according to \eqref{deltaLambdaPos} and still end up with a negative unstable potential as in Figure \ref{Wrong}.
\end{minipage}

To avoid this kind of behaviour, one can already impose that the large field values of the potential and its derivative be positive:
\begin{equation}
	\lim_{\sqrt{\xi}\chi\to \infty}U(\chi) = \frac{1}{4\xi^2}\left(\Bar{\lambda}_0 + b\log^2\left(\frac{y_t}{q\sqrt{\xi}}\right)\right) >0\;,
	\label{PotAsympt}
\end{equation}
\begin{equation}
	\lim_{\sqrt{\xi}\chi\to \infty}\frac{d U}{d\chi} = 
	\frac{1}{\xi^{3/2}}\frac{\tanh^3(\sqrt{\xi}\chi)}{\cosh^2(\sqrt{\xi}\chi)}\left(\Bar{\lambda}_0 + \frac{b}{2} \log\left(\frac{y_t}{q\sqrt{\xi}}\right)
	+ b \log^2\left(\frac{y_t }{q\sqrt{\xi}}\right)\right)>0\;.
	\label{DerPotAsympt}
\end{equation}
This can be rewritten as constraints in the $(\xi,\lambda_0,\delta\lambda)$ space. First, the constraint \eqref{PotAsympt} is trivial for $\Bar{\lambda}_0>0$. Otherwise, it translates to the following condition on $\xi$:
\begin{equation}
	\xi \leq \left(\frac{y_t}{q}\right)^2 e^{-2\sqrt{\frac{|\Bar{\lambda}_0|}{b}}}\text{ for $\Bar{\lambda}_0<0$ and \eqref{PotAsympt}}\;.
	\label{XiAsympt}
\end{equation}
Then, equation \eqref{DerPotAsympt} translates to
\begin{equation}
	\xi \leq \left(\frac{y_t}{q}\right)^2 e^{-\frac{1}{2}\left(-1+\sqrt{1-\frac{16\Bar{\lambda}_0}{b}}\right)}\text{ for \eqref{DerPotAsympt}}\;.
	\label{DerXiAsympt}
\end{equation}
Notice that both \eqref{PotAsympt} and \eqref{DerPotAsympt} are automatically satisfied for $\Bar{\lambda}_0 > \frac{b}{16}$.

Now, let us determine which of these conditions to use when $\Bar{\lambda}_0<0$. In this overlapping region, it can be shown that the bound in \eqref{XiAsympt} is systematically lower, and one does not need the condition \eqref{DerXiAsympt}.

In summary, the asymptotic features of the potential are correctly restored in the jump case where:
\begin{equation}
	\left\{
	\begin{array}{crc}
		\xi & \leq & \left(\frac{y_t}{q}\right)^2 e^{-2\sqrt{\frac{|\Bar{\lambda}_0|}{b}}} \text{ for $\Bar{\lambda}_0 < 0$}\\
		\xi & \leq & \left(\frac{y_t}{q}\right)^2 e^{-\frac{1}{2}\left(-1+\sqrt{1-\frac{16\Bar{\lambda}_0}{b}}\right)}\text{ for $\Bar{\lambda}_0<\frac{b}{16}$}\\
		\xi & \leq & \xi_{\text{thresh}}\text{ Otherwise}\;.
	\end{array}
	\right.
	\label{CondAsympt}
\end{equation}
\begin{enumerate}
	\item $\boldsymbol{\lambda_0<0}$.
	
In this case, the jump may actually be necessary if one wants to ensure the stability of the potential at the feature. Indeed, by making use of the conditions \eqref{CondAsympt}, one can show that there is a minimal value of the jump parameter such that the potential is still positive at large field values:
	\begin{equation}
		\delta\lambda > \delta\lambda_{\text{min}} = |\lambda_0|-\log^2\left(\frac{y_t }{q\sqrt{\xi}}\right)\sim 
		|\lambda_0| - 0.25\log^2(\xi) +0.3\log(\xi) -0.09\;,
		\label{DelMin}
	\end{equation}
	where we used \eqref{ParamSpace} for the estimations. This is maximal for $\xi = (y_t/q)^2\sim 1.8$.\\
	If $|\lambda_0| < b\log^2(y_t/q\sqrt{\xi})< 10^{-5}$, there is no jump needed to save the potential at large field values. The local minimum feature is therefore completely provided by the jumpless physics, and the positive jump would only drive us away from there. In what follows, we will therefore implicitly assume that the absolute value of $\lambda_0$ is larger than $\mathcal{O}(10^{-5})$.
	Now, let us briefly discuss the allowed space for having a critical point in the potential. Because of the negative $\lambda_0$, the minimum of the running is necessarily negative. Then, the jump drives it into a positive value, which can only result in a local maximum. The local minimum can only arise from the little bump in $G^2(\chi)$ at $\chi = \chi_2$ \eqref{LittleBump}. Because of the universality of $G^2$'s properties, the depth of the local minimum only depends on $\delta\lambda$. Consequently, the resulting local minimum may reach a negative value if the jump is too small. This introduces the need for a lower bound $\delta_->0$ on the deviation from the bound \eqref{DelMin}.
	
	 On the other hand, it may be diluted by the jump if the latter has a too large magnitude. The intuition behind this is that the jump parameter value $\delta\lambda$ cannot be too far away from the bound \eqref{DelMin}, introducing a $\delta_+$ such that:
	\begin{equation}
		\delta\lambda(\xi) \in [\delta\lambda_{min}(\xi)+\delta_-,\delta\lambda_{\min}(\xi)+\delta_+]\;.
	\end{equation}
	\begin{minipage}{0.45\linewidth}
		\begin{figure}[H]
			\hspace{-0.9cm}
			\includegraphics[scale = 0.6]{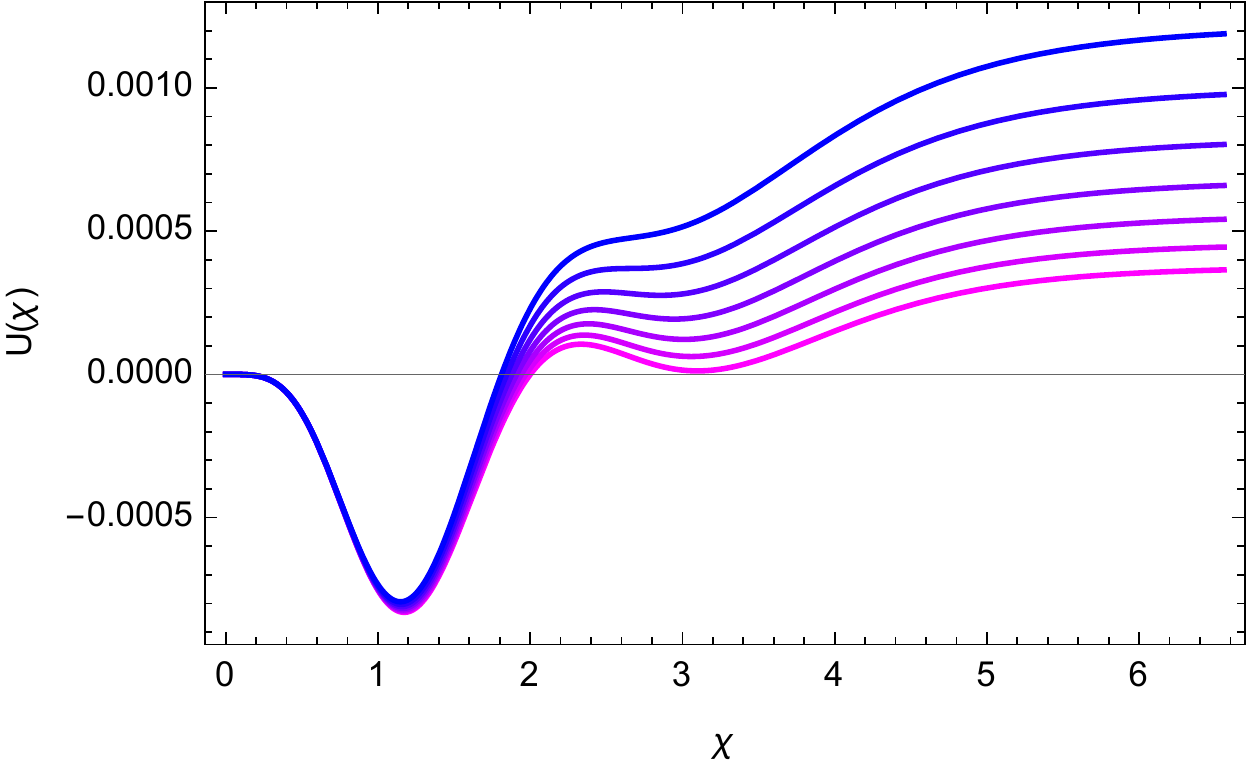}
			\caption{Visualisation of the effect of the jump in the potential. The blue region corresponds to $\delta\sim\delta_+$ and the pink one stands for $\delta\sim\delta_-$.}
		\end{figure}
	\end{minipage}
	\hfill
	\begin{minipage}{0.45\linewidth}
		To quantify this, one has to numerically scan over the possible jump parameters and identify the allowed region. If $\xi$ or $|\lambda_0|$ are too small, one can observe that one reaches $\delta_+$ before $\delta_-$. In this case, the bound $\delta_-$ does not exist and the potential cannot be tuned to the correct shape. In practice, the limiting value of $\xi$ appears to be quite close to the threshold value $\xi_{thresh}$ and $\lambda_0$ greater than a bound which can be empirically found numerically.
	\end{minipage}
	
	The relevant parameter space for a negative coupling is then:
	\begin{equation}
		\xi \lesssim \xi_{thresh},\quad \delta\lambda\gtrsim \delta\lambda_{min}(\xi),\quad \lambda_0 \lesssim \mathcal{O}(-5\times10^{-4})\;.
		\label{ParamNeg}
	\end{equation}
	
	\item $\boldsymbol{\lambda_0 > 0}, \boldsymbol{\delta\lambda>0}$
	
	In this region, one can either consider a negative or a positive jump parameter $\delta\lambda$. As we already stated, one can always tune this parameter in order to get a local minimum by making use of equation \eqref{deltaLambdaPos}. Let us consider the case where $\delta\lambda>0$. In this case, one does not expect non-trivial behaviour due to the jump. However, one can see that there is still a small parameter space region allowed by equation \eqref{deltaLambdaPos}. To see this, let us recall that $A(\chi)-1<1$ for any $\chi$. Thus, the jump can be made positive in the region of field space where the numerator of \eqref{deltaLambdaPos} is negative. This 
	can be achieved between $\chi_-$ and $\chi_+$ defined in \eqref{PalatniChiPm} in the region where $\Bar{\lambda}_0< b/16$.
	
	Since we are considering a positive jump here, this is therefore a restriction of the jumpless region. Indeed, one can write that $\lambda_0 < b/16-\delta\lambda <b/16$. The positive jump requirement also implies that $\delta\lambda$ cannot be larger than $b/16$. By looking at the asymptotic conditions \eqref{CondAsympt}, one can also derive an upper bound on $\xi$ which can be thought of as the same as in the jumpless case.
	
	\item $\boldsymbol{\lambda_0 > 0}, \boldsymbol{\delta\lambda<0}$
	
	Elsewhere, the jump is a negative quantity. Let's examine what happens. As before, we should restrict the parameter space in order to keep the potential positive at large field values. Thus, \eqref{CondAsympt} gives us the smallest possible value for the jump:
	\begin{equation}
		0>\delta\lambda>\delta\lambda_{min}(\xi) = -\left(\lambda_0 + b\log^2\left(\frac{y_t}{q\sqrt{\xi}}\right)\right)\;.
	\end{equation}
	Now, one can adopt the same viewpoint as before and parametrise the jump from this minimal value thanks to a parameter $\delta>0$. As before, $\delta$ can reach a maximal value $\delta_+$ where all the jump features disappear from the potential. Analytically, this corresponds to the jump parameter coming close to zero and becoming negligible in the description. One is then reducing to the jumpless situation.
	
	\begin{minipage}{0.45\linewidth}
		\begin{figure}[H]
			\hspace{-0.9cm}
			\includegraphics[scale = 0.6]{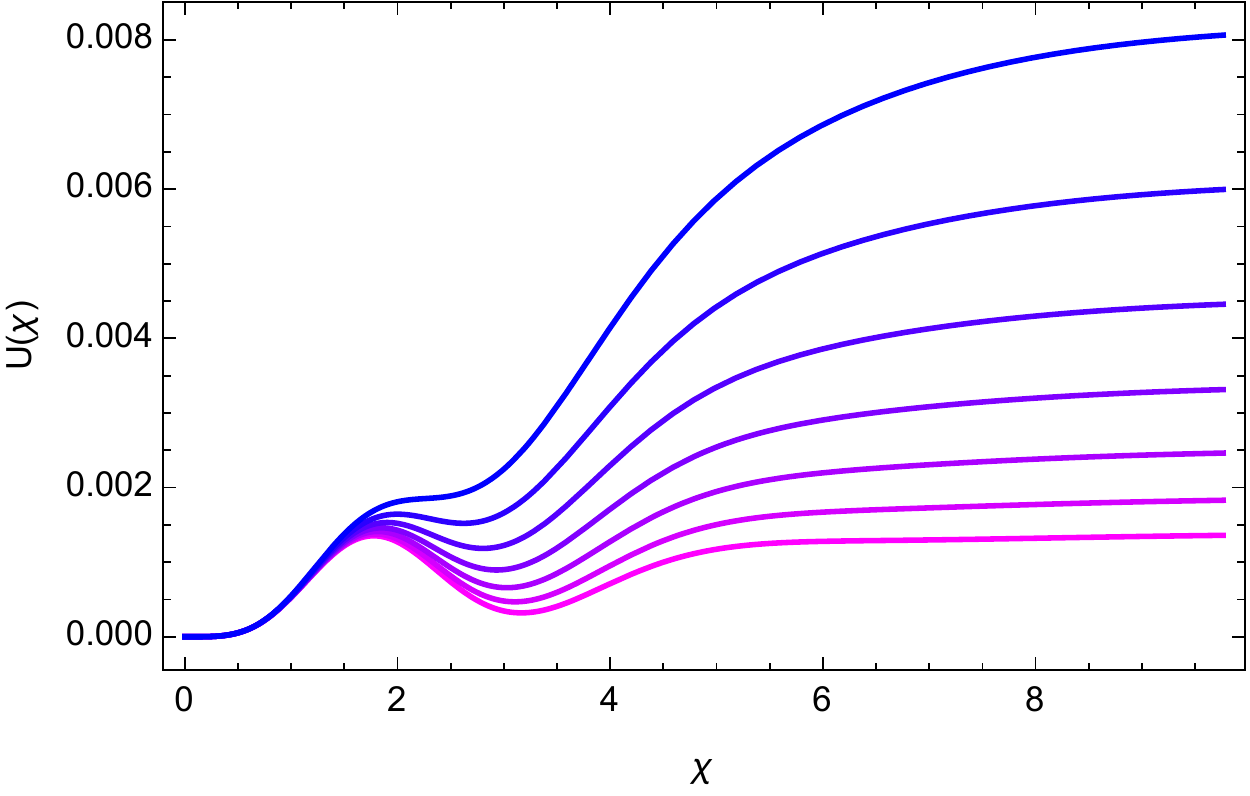}
			\caption{Visualisation of the effect of the jump in the potential. The blue region corresponds to $\delta\sim\delta_+$ and the pink one stands for $\delta\sim\delta_-$.}
		\end{figure}
	\end{minipage}
	\hfill
	\begin{minipage}{0.45\linewidth}
		Next, let's see if a lower bound can be imposed. For $\delta<\delta_+$, the potential begins to grow because of the positive $\lambda_0$. Then, there is the jump and it goes down to a local minimum. Following this, the growing nature of the potential dominates again and we reach the usual asymptotic behaviour. Now, we have to impose two things: the local maximum at the jump may not be a global maximum and the local minimum has to be positive. As in the negative case, we do not take the too small $\lambda_0$ into account since they are reducing to the jumpless case.
	\end{minipage}
\end{enumerate}

\subsection{Numerical Scan Over the $(\xi,\delta\lambda)$ Space}
\label{ssec:jumpParameterSpace}
\begin{figure}[H]
	\begin{subfigure}{.45\textwidth}
		\centering
		\includegraphics[width=1.0\linewidth]{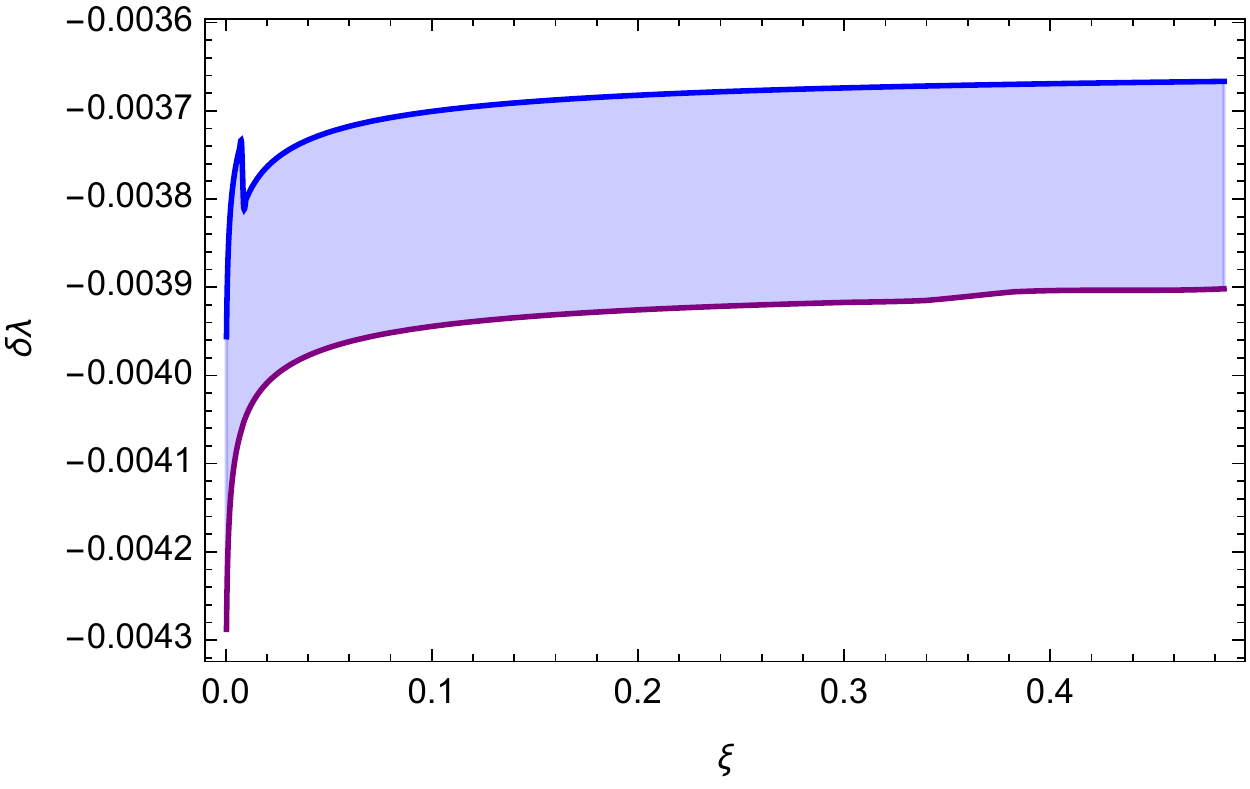}
		\caption{Allowed Parameter Space for $\lambda_0= 0.004$ and $y_t^{low} = 0.918$}
	\end{subfigure}
	\hspace{2em}
	\begin{subfigure}{.45\textwidth}
		\centering
		\includegraphics[width=1.0\linewidth]{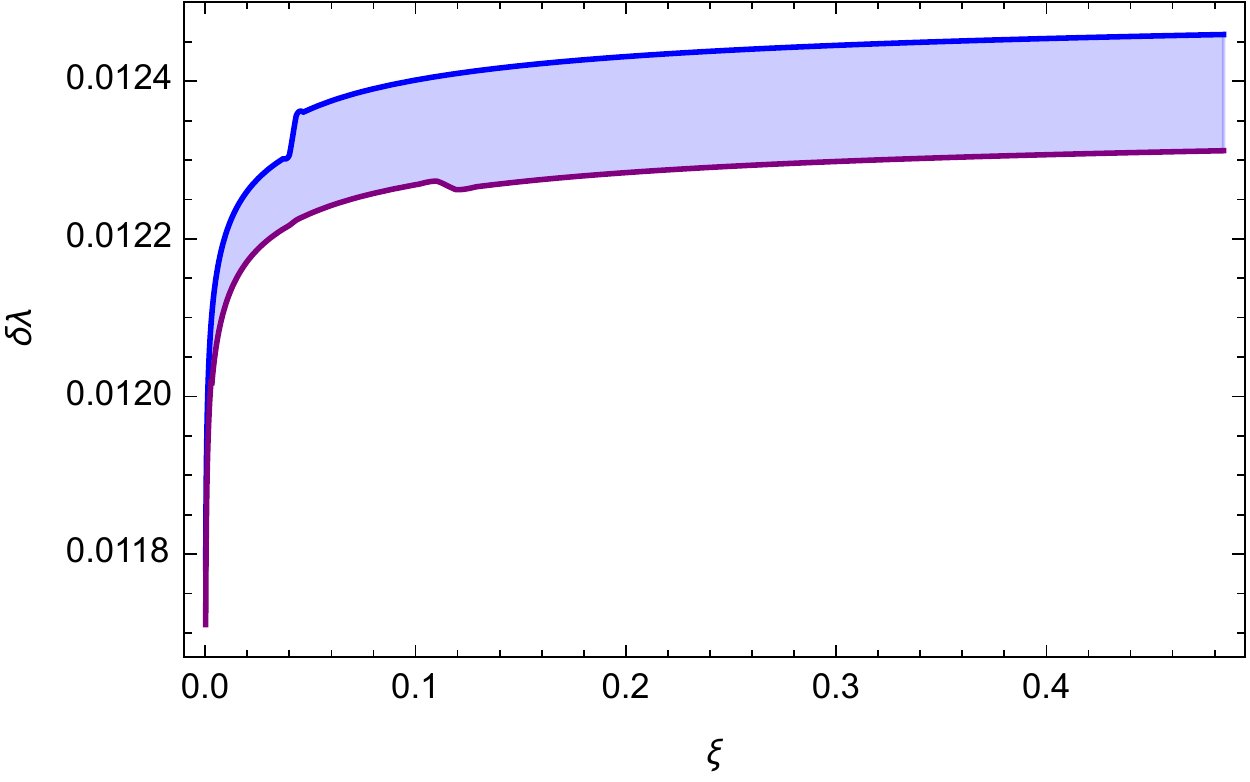}
		\caption{Allowed Parameter Space for $\lambda_0=- 0.01$ and $y_t^{low} = 0.934$}
	\end{subfigure}
	\caption{Regions in the $(\xi,\delta\lambda)$ space which are consistent with a local minimum in the inflationary potential for two values of the top Yukawa.}
	\label{BlueRegions}
\end{figure}

The following part provides a concrete visualisation of the allowed parameter space in order to obtain a stable potential with a local minimum in the jumping case. The procedure described above allows us to systematically find the bound $\delta_\pm$ for a given value of $\xi$ and $y^{low}_t$. The next thing to do is to scan over all possible $\xi$ at given values of the top mass in order to get an idea of the possible jump value. The result of these scans is showed in figure \ref{BlueRegions} for two values of the top mass corresponding to positive (left) and negative (right) $\lambda_0$. The regions which are compatible with a critical point are shown in blue.

For $\lambda_0< b/16$, one can observe a merging between the blue region and the $y$ axis due to the intersection with the allowed jumpless situations. The system is then effectively reduced to the latter.

\section{Incompatibility with CMB Normalisation} \label{sec:CMB}
In our framework, quantum effects leading to a significant enhancement of the generated amplitude of perturbations can only occur during inflation if $\xi \lesssim 1$ (see eqs.~\eqref{xiMaxInflection} and \eqref{xiMaxJump}). In contrast, the tree-level analysis of inflation leading up to eq.~\eqref{normalizationCMBTree} shows that matching the observed CMB normalisation requires a very large $\xi \sim 10^7$. This already provides an indication that $\xi \lesssim 1$ is incompatible with the observed amplitude of CMB perturbations, even beyond tree-level analysis.

Before proceeding to exact numerical analysis, we shall present analytic arguments suggesting that this is indeed the case. To this end, we will first display the tree-level result of CMB normalisation in the form (\cf \eqref{normalizationCMB})
\begin{equation}
	\frac{U^{\text{SR}}_{\text{CMB}}}{\epsilon^{\text{SR}}_{\text{CMB}}} \approx \frac{\lambda}{128\xi^3} \sinh^2\left(2\sqrt{\xi} \chi^{\text{SR}}_{\text{CMB}}\right)  \tanh^4\left(\sqrt{\xi} 	\chi^{\text{SR}}_{\text{CMB}}\right) \;,
\end{equation}
where we used eqs.~\eqref{PotF}, \eqref{FPalatini}, and \eqref{EpsPalTree}.
 Now we can ponder how this analysis changes once quantum effects are included. The key point is as follows. Whereas the possible presence of an inflection point or a jump of the coupling constant leads to strong deviations from slow-roll dynamics locally, these features are generically far away from the field value $\chi_{\text{CMB}}$, where CMB perturbations are produced. As long as slow roll is a good approximation in the vicinity of $\chi_{\text{CMB}}$, we can still employ \eq \eqref{normalizationCMB}, after taking into account two effects. First, strong deviations from tree-level analysis and slow-roll behaviour can occur between the generation of CMB perturbations and the end of inflation. Therefore, \eq \eqref{eqN} is not valid any more -- not even approximately -- and it becomes impossible to infer the field value $\chi_{\text{CMB}}$ from simple analytic arguments. Second, the value of $\lambda$ that is relevant at inflationary energies can change due to the effects of RG running. This leads us to the following approximation for the CMB normalisation
\begin{equation} \label{normalizationCMBApproximate1}
	\frac{U_{\text{CMB}}}{\epsilon_{\text{CMB}}} \approx \frac{\lambda(\chi_{\text{CMB}})}{128\xi^3} \sinh^2\left(2\sqrt{\xi} \chi_{\text{CMB}}\right)  \tanh^4\left(\sqrt{\xi} 	\chi_{\text{CMB}}\right) \;,
\end{equation}
where, as shown in \eq \eqref{FullPot}
\begin{equation}
	\lambda(\chi) =	\left(\lambda_0 + b \log^2\left(\frac{y_t F(\chi)}{q}\right) + \delta\lambda\Big(1-G^2(\chi)\Big)  \right) \;.
\end{equation}

For the analytic estimate, we shall consider a situation in which $\chi_{\text{CMB}}$ is not much smaller than its tree-level value \eqref{chiCMB},
\begin{equation} \label{chiCMBApproximate}
	\chi_{\text{CMB}} \gtrsim	\chi^{\text{SR}}_{\text{CMB}} = \frac{\arccosh\Big(16 \xi N_\star+\sqrt{32\xi + 1}\Big)}{2 \sqrt{\xi}} \;.
\end{equation}
Moreover, we shall assume that $\xi \gtrsim 1/N_\star$. As will become evident later, most parts of parameter space fulfill this mild condition.\footnote
{For completeness, we discuss the opposite case of vanishing $\xi$ in appendix \ref{app:SmallXi}. Also in this case, we find incompatibility with the amplitude of CMB perturbations.}
Then \eq \eqref{chiCMB} corresponds to
\begin{equation} \label{chiBound}
	\chi_{\text{CMB}}  \gtrsim \frac{3.5}{2 \sqrt{\xi}} \;.
\end{equation}
Thus, it follows from eq \eqref{FCMB} that 
\begin{equation}
	F^{\text{SR}}_{\text{CMB}} \gtrsim \frac{1}{\sqrt{\xi}} \tanh\left(\frac{3.5}{2} \right) \gtrsim \frac{0.94}{\sqrt{\xi}} \;,
\end{equation}
\ie CMB perturbations are generated on the asymptotic plateau of the potential. In this situation, \eq \eqref{normalizationCMBApproximate1} is bounded:
\begin{equation}
	\frac{U_{\text{CMB}}}{\epsilon_{\text{CMB}}} \gtrsim \frac{\lambda(\chi_{\text{CMB}})}{\xi^3}\;.
\end{equation}
Thus, the observed normalisation of perturbations in the CMB, $\frac{U_{\text{CMB}}}{\epsilon_{\text{CMB}}} = 5 \cdot 10^{-7}$, cannot be obtained unless the Higgs self-coupling assumes very small values $\lambda(\chi_{\text{CMB}}) \lesssim 10^{-7}$.\footnote
{It is important to emphasise that a tiny $\lambda(\chi_{\text{CMB}})$ is a necessary but not a sufficient condition. Regardless of the value of $\lambda(\chi_{\text{CMB}})$, we have not been able to find a single example in which $\xi<1$ can lead to successful inflation matching the observed CMB normalisation.}

Finally, we can go one step further and obtain a more accurate approximation than that shown in eq \eqref{chiCMBApproximate}. We recall that in deriving eq \eqref{chiCMBApproximate}, we neglected all derivatives of $\lambda(\mu)$. It is straightforward to include them, and then we obtain from eq \eqref{Derivative}:
\begin{equation} \label{normalizationCMBApproximate2}
	\frac{U_{\text{CMB}}}{\epsilon_{\text{CMB}}} \approx \mathcal{D}(\chi_{\text{CMB}}) \frac{\lambda(\chi_{\text{CMB}})}{128\xi^3} \sinh^2\left(2\sqrt{\xi} \chi_{\text{CMB}}\right)  \tanh^4\left(\sqrt{\xi} 	\chi_{\text{CMB}}\right) \;,
\end{equation}
where
\begin{equation}
	\mathcal{D}(\chi)= \frac{\lambda^2(\chi)}{\left(\lambda(\chi) + \frac{b}{2}\log\left(\frac{y_t F(\chi)}{q}\right)-\delta \lambda \frac{G'(\chi)G(\chi)F(\chi)}{2F'(\chi)}\right)^2} \;.
\end{equation}
Both additional terms in the denominator cannot lead to a significant suppression as long as $\lambda(\chi_{\text{CMB}}) \gtrsim 10^{-7}$. The first contribution is of the order of $b\sim 10^{-5}$, and the second one is suppressed by $G'(\chi)$, which vanishes for large $\chi$. We expect that \eq \eqref{normalizationCMBApproximate2} yields a viable approximation as long as slow roll is valid during CMB generation. One can expect that a violation of slow roll does not improve the situation since generically this leads to an enhancement of perturbations, as discussed in section \ref{ssec:USR}.
These heuristic arguments will be confirmed by the subsequent numerical analysis.

\section{Full numerical analysis}\label{sec:Numerics}

In the previous sections, we explicitly found the necessary parameter space to produce a local minimum in the inflationary potential, and we provided simple analytical arguments which points toward an incompatibility with the observed CMB normalization \eqref{PLANCK}. The next step is then to numerically scan over the allowed regions of the $(\lambda_0,\xi,\delta\lambda)$ space. For each relevant points, we compute the primordial power spectrum of the comoving curvature perturbation $\mathcal{R}$. As stated in section \ref{sec:GenInfl}, the standard way to proceed is to first solve for the classical background dynamics \eqref{Friedman} and then to solve the Mukhanov-Sasaki equation \eqref{MukhanovConf} for each relevant setting Bunch-Davis initial condition. To do this numerically, we translate the dynamical equations in terms of the number of e-folding $\diff N = H \diff t$.

First, all the background equations can be combined to give:
\begin{equation}
	\frac{\diff^2\chi}{\diff N^2} + 3\frac{\diff\chi}{d N} - \frac{1}{2}\left(\frac{\diff\chi}{\diff N}\right)^3 + \left(3-\frac{1}{2}\left(\frac{\diff\chi}{\diff N}\right)^2\right)\frac{U'(\chi)}{U(\chi)} = 0\;.
	\label{eom}
\end{equation}
To solve this, we take the initial condition $\chi_i$ to be far away in the asymptotic plateau of $U(\chi)$ and we choose the initial velocity to be $\chi'(N_i) = -\sqrt{2\epsilon^{SR}(\chi_i))}$. Then, we evolve until we reach a point where $\epsilon_H(N_{end})=1$ which defines the endpoint. The CMB generation is then identified with $N_{CMB} = N_{end}-55$.

The second step is then to solve the Mukhanov-Sasaki equation \cite{Ballesteros:2017fsr,Ballesteros:2020qam}:
\begin{equation}
	\frac{\diff^2 f_{k}}{\diff N^2} + (1-\epsilon_H)\frac{\diff f_k}{\diff N} + \left(\frac{k^2}{a^2H^2}+ (1+\epsilon_H-\eta_H)(\eta_H-2)-\frac{\diff}{\diff N}\left(\epsilon_H-\eta_H\right)\right)f_k = 0\;.
	\label{Mukhanov}
\end{equation}
Then, to set the Bunch-Davies initial condition \eqref{BunchDavis}, it is in general sufficient to take an initial time $N_i$ such that the relevant mode $k$ for the computation is still well inside the horizon. Namely, 
\begin{equation}
	k\gg H(N_i) a(N_i) = k_i\;,
\end{equation}
where $k_i$ is the mode that crossed the horizon at $N_i$. By switching the variable $\tau$ for $N$ and the initial condition \eqref{BunchDavis} becomes:
\begin{equation}
	f_k(N_i) = \frac{1}{\sqrt{2 k}} \,,\quad f_k'(N_i) = - \frac{i}{k_i} \sqrt{\frac{k}{2}} \;.
\end{equation}
Usually, choosing $k_i$ one hundred or one thousand smaller than $k$ is sufficient for our purpose.

\subsection{Scenario without jump}
As above, let us start our discussion with the simple scenario where the theory does not feature any jump in the coupling constant coming from a possible UV completion. In that case, the theory only depends on a two-dimensional parameter space $(y^{low}_t,\xi)$. Although this might be a large space to deal with, let us recall that we already derived the necessary conditions \eqref{Const1} and \eqref{Const2} for the potential to feature a local minimum.

The strength of the enhancement mainly depends on the time the background field spend in the USR phase. Let us see what happens at some fixed top mass. Let us notice that the depth of the local minimum does not depend on $\xi$:
\begin{equation}
	U(\chi_\pm) = \frac{1}{4\xi_{max}^2}\left(\lambda_0 + b\log^2(E_\pm)\right)\;,
\end{equation}
where $\xi_{max}$ is the upper bound on the non-minimal coupling that we derived on section \ref{ParamSpa} (see equation \eqref{xiMaxInflection}). On the other hand, the height of the asymptotic plateau is suppressed by $1/\xi^2$:
\begin{equation}
	U\left(\chi\gg 1/\sqrt{\xi}\right)\to \frac{1}{4\xi^2}\left(\lambda_0 + b\log^2\left(\frac{y_t}{q\sqrt{\xi}}\right)\right)\;.
\end{equation}
Thus, for a fixed value of the potential barrier formed by the local maximum at $\chi_-$, the height of the field's starting point may decrease as $\xi$ increases. Thus, the incoming kinetic energy of the field at the local minimum may be smaller as $\xi$ greater. However, one cannot necessarily set $\xi=\xi_{max}$ to get a large enhancement. Indeed, there may be a limit value $\xi_{lim}<\xi_{max}$ such that the field fall in the local minimum without a sufficient amount of kinetic energy to overcome the barrier.

\begin{figure}[H]
	\begin{subfigure}{.5\textwidth}
		\centering
		\includegraphics[width=.9\linewidth]{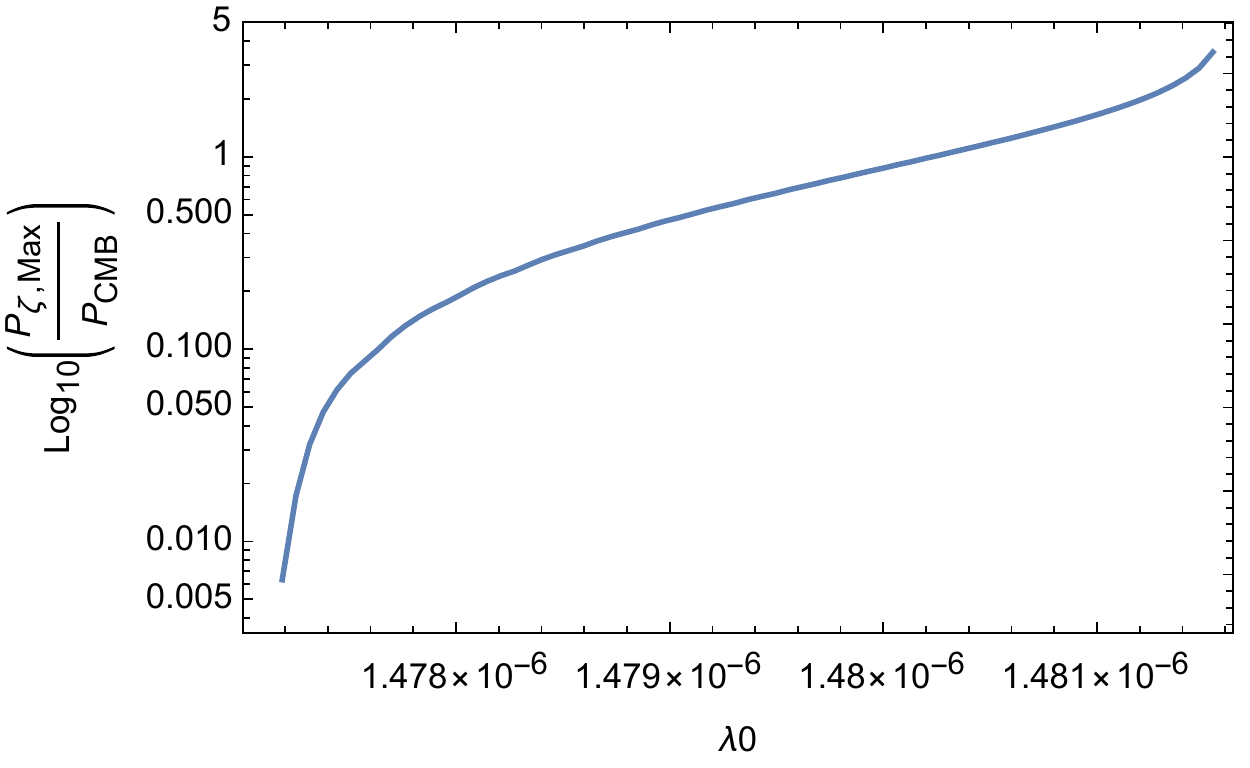}
	\end{subfigure}
	\begin{subfigure}{.5\textwidth}
		\centering
		\includegraphics[width=.9\linewidth]{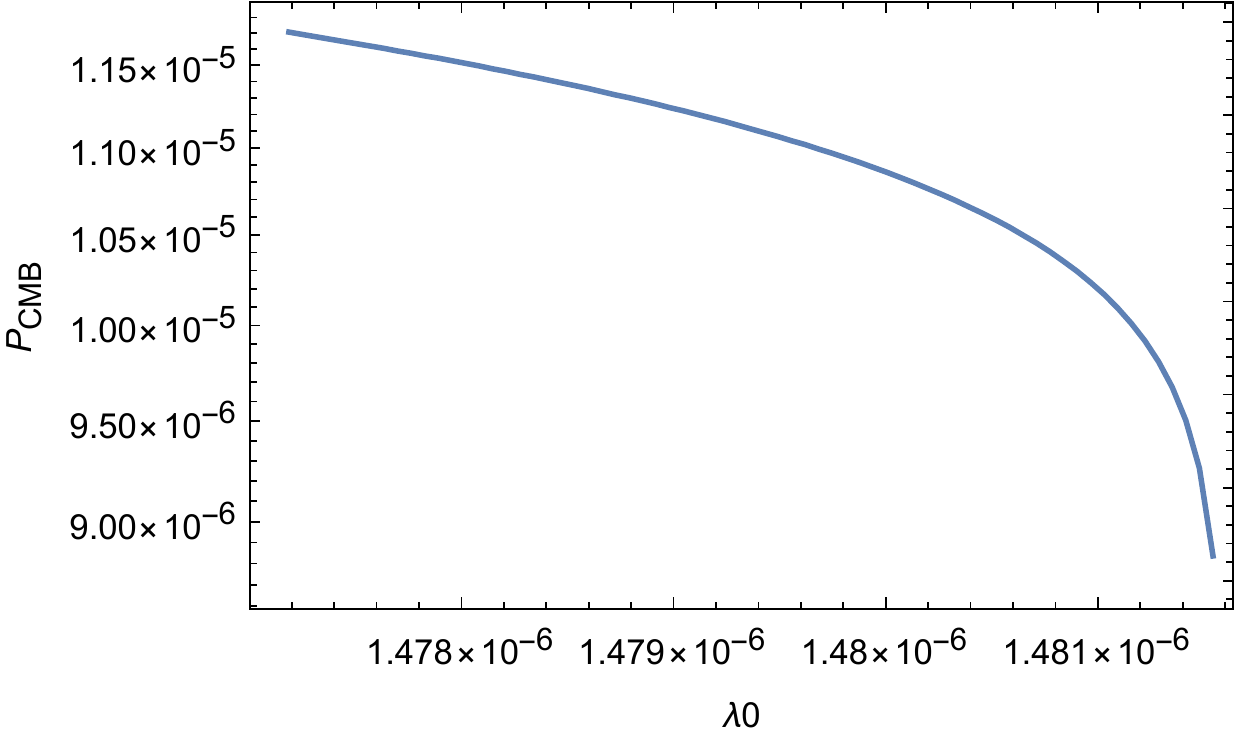}
	\end{subfigure}
	\begin{subfigure}{.5\textwidth}
		\centering
		\includegraphics[width=.9\linewidth]{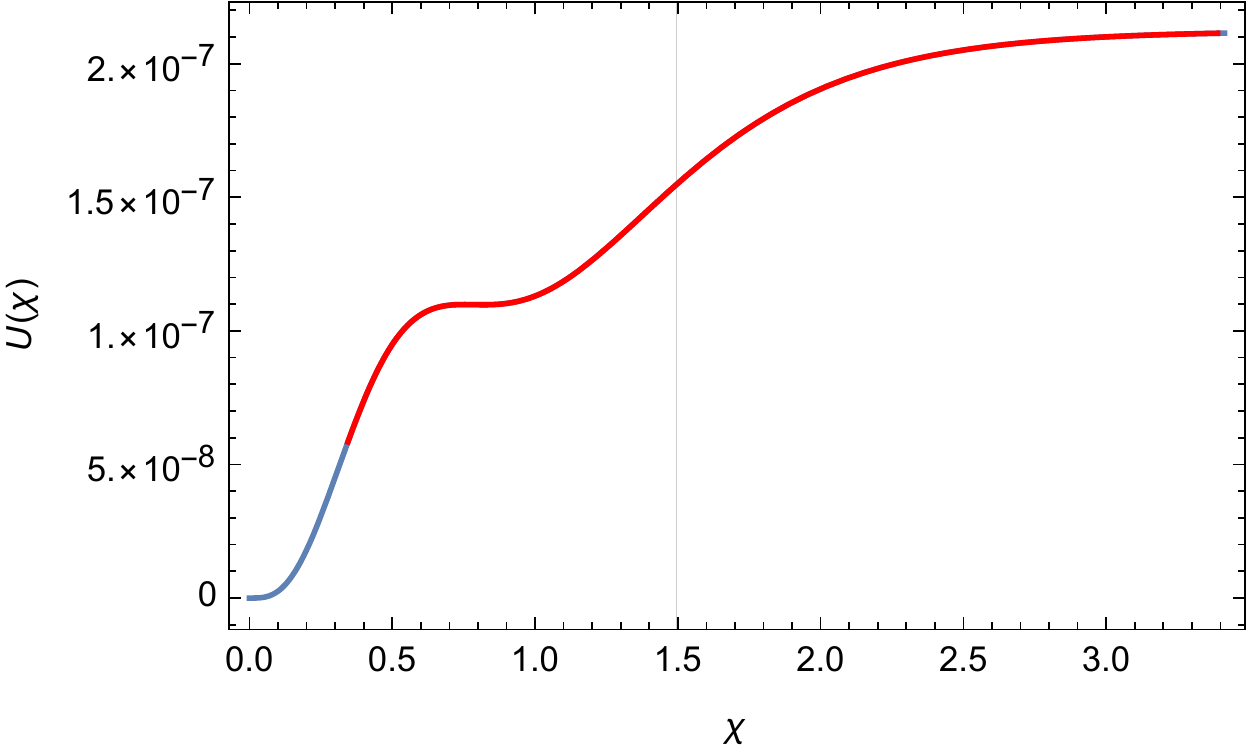}
	\end{subfigure}
	\begin{subfigure}{.5\textwidth}
		\centering
		\includegraphics[width=.9\linewidth]{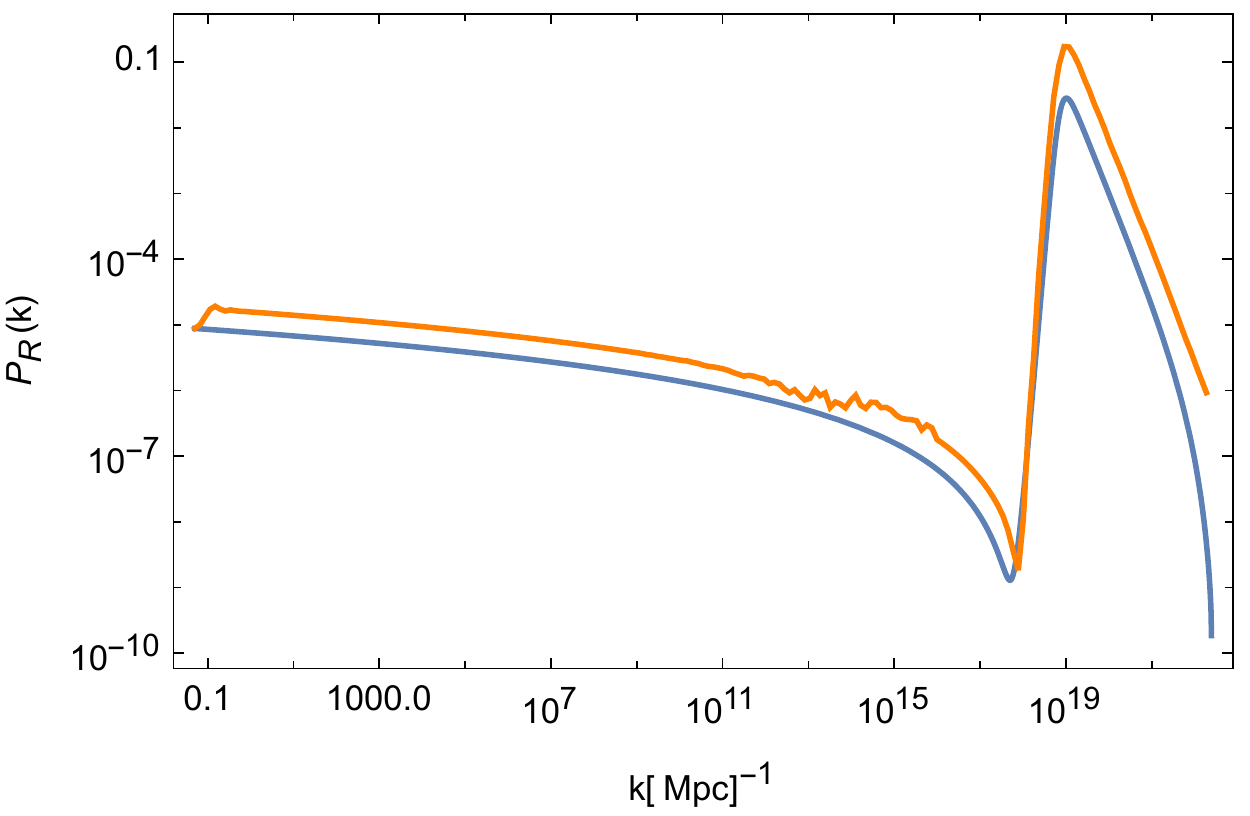}
	\end{subfigure}
	\caption{\textbf{Upper Left}: Maximal Enhancement of the scalar power spectrum as a function of the parameter $\lambda_0$. This corresponds to a very narrow window of top Yukawas around $y_t^{low}\sim 0.923$ (roughly $m_t^{pole}\sim 170$GeV). One can see that the optimal enhancement happens for the largest value of $\lambda_0$. \textbf{Upper Right}: Value of the power spectrum at the CMB normalization. This reaches its minimal value at the largest value of the enhancement. However, it is still (at least) four orders of magnitude above the observational constraints. \textbf{Lower Left}: Inflationary trajectory (in red) along the inflationary potential (in blue) for the optimal enhancement scenario. The vertical bar in the plot is at the location of the lower bound on the initial condition given by equation \eqref{chiBound}. \textbf{Lower Right}: Scalar power spectrum as function of the scale for the optimal enhancement scenario. The blue curve shows the results obtained using the approximate equation \eqref{PowSR}) and the orange one comes from solving the exact Mukhanov-Sasaki equation \eqref{Mukhanov}.}
	\label{ResultsJpLess}
\end{figure}

Because of the non-linearity of the equation of motion, one has to find this $\xi_{lim}$ numerically for each possible value of $y^{low}_t$. The result of a scan over the (tiny) allowed region is shown on figure \ref{ResultsJpLess}. The corresponding window of the top mass being very small, the more transparent thing to do was to parametrised the scan result as function of the minimum of the quartic coupling's running $\lambda_0$.
After making the numerical scan, one gets to the conclusion that the greater enhancement corresponds to the minimal value for the CMB normalization. This corresponds to $\lambda_0 = 1.482\cdot 10^{-6}$ and $\xi_{lim} = 1.373$ when $\xi_{max}(\lambda_0) = 2.353> \xi_{lim}$. The full inflationary dynamics of this system is shown on the two lower plots of figure \ref{ResultsJpLess}. The power spectrum at the CMB is way greater than the PLANCK constraints \eqref{PLANCK}.
Notice that equation \eqref{Mukhanov} and \eqref{PowSR} agree on this value with a very nice precision.

\begin{figure}[H]
	\begin{subfigure}{.5\textwidth}
		\centering
		\includegraphics[width=.9\linewidth]{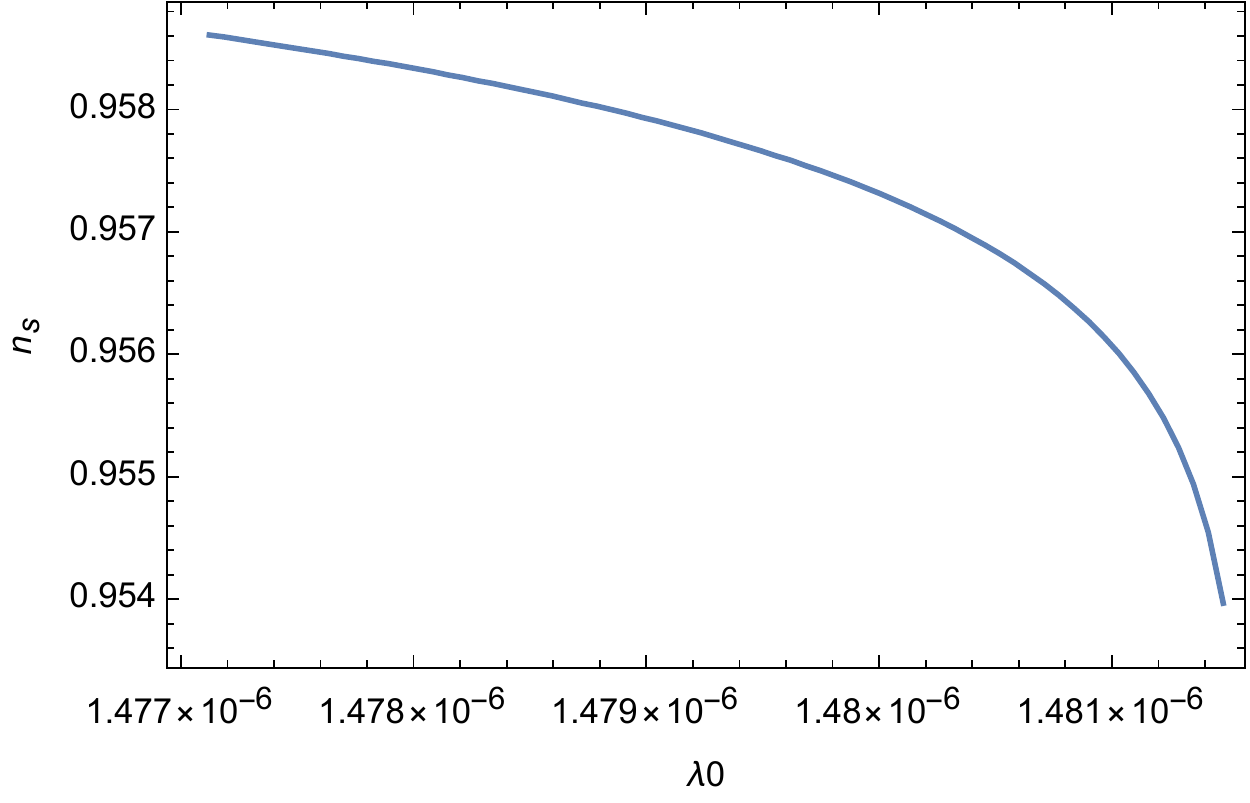}
	\end{subfigure}
	\begin{subfigure}{.5\textwidth}
		\centering
		\includegraphics[width=.9\linewidth]{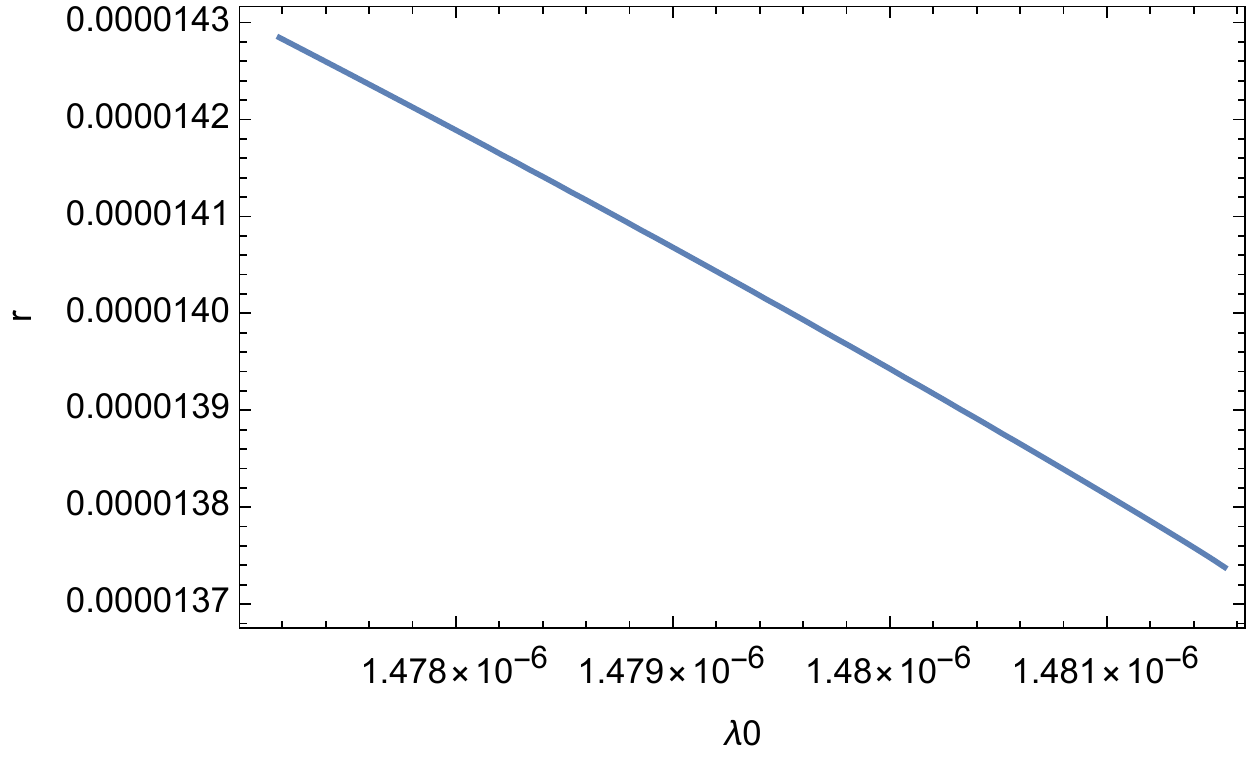}
	\end{subfigure}
	\caption{Inflationary observables as function of $\lambda_0$ for $\xi = \xi_{lim}(\lambda_0)$. \textbf{Left}: Plot of the CMB spectral tilt. \textbf{Right}: Plot of the tensor to scalar ratio.}
	\label{ObsJpLess}
\end{figure}

To be complete, let us mention the fact that the other standard inflationary observables behave a bit better. Figure \ref{ObsJpLess} shows the results of the scan for the tensor to scalar ratio and the spectral index at the CMB scale. They are closer the observational constraints \eqref{nsPLANCK} and \eqref{rPLANCK}, although there still is a tension in the spectral index. To conclude, we saw that the parameter space that produces a large enhancement of the scalar power spectrum at small scales is not compatible with the Palatini Higgs inflation without taking any UV completion effect into account.

\subsection{Scenario with jump}

In this section, we will reproduce the same kind of analysis with a non-vanishing jump. As explained above, the parameter space is then promoted to a three-dimensional space $(y^{low}_t,\xi,\delta\lambda)$ under which we already put some necessary constraints in order to see a local minimum in the Einstein frame potential. As we already explained, the complex form of the function $G(\chi)$ \eqref{GConcrete} that materialize the actual jump in the quartic coupling forbids a full analytical treatment of the master equation \eqref{MasterEq}. Consequently, one has to rely on the numerical procedure defined in section \ref{ParamSpa} to find the allowed subspace in $(y_t^{low},\xi,\delta\lambda)$. As before, the goal will be to ultimately to provide the same kind of information than in figure \ref{ResultsJpLess}: we need an optimization procedure at fixed value of $y^{low}_t$ to find the maximal enhancement and the minimal CMB. We considered a range of top Yukawas between $y_t^{low}\in[0.918,0.941]$ which roughly corresponds to $m_t^{pole}\in[170,174]$ GeV (see \eg \cite{Bezrukov:2014ina, Shaposhnikov:2020fdv} for discussions regarding observationally viable ranges of $y_t^{low}$).

First, equation \eqref{xiMaxJump} shows that the requirement to see a non-trivial jump effect introduces a threshold value of $\xi$ such that $\xi<\xi_{threshold}\sim 0.47$. For larger $\xi$, the effect of the jump results in a constant shift in the quartic coupling and the system effectively reduces to the jumpless scenario where the jump effect is fully used to satisfy the constraint \eqref{Const1} and $\xi$ must satisfy equation \eqref{Const2}. Such cases have already been ruled out by the previous section on the jumpless theory. Moreover, we empirically realized that there were a lower bound on $\xi$ under which inflation does not end, and the field get stuck in the local minimum.

Then, for some given $\xi$, the value of $\delta\lambda$ is restricted to a very small interval $\delta\lambda\in [\delta\lambda_-,\delta\lambda_+]$ which can be found numerically using the procedure presented in section \ref{ParamSpa}. As we already mentioned several times above, the value of the enhancement strongly depends on the time the background field spent into the local minimum. Moreover, we showed in section \ref{ParamSpa} that the size of the local minimum is directly related to the magnitude of $\delta\lambda$: the maximal depth is reached at $\delta\lambda_-$ and the local minimum simply disappears above $\delta\lambda_+$. The procedure is then to progressively decrease the value of the jump until we reached the minimal value such that field does not get stuck into the local minimum.

\begin{figure}[H]
	\begin{subfigure}{.5\textwidth}
		\centering
		\includegraphics[width=.9\linewidth]{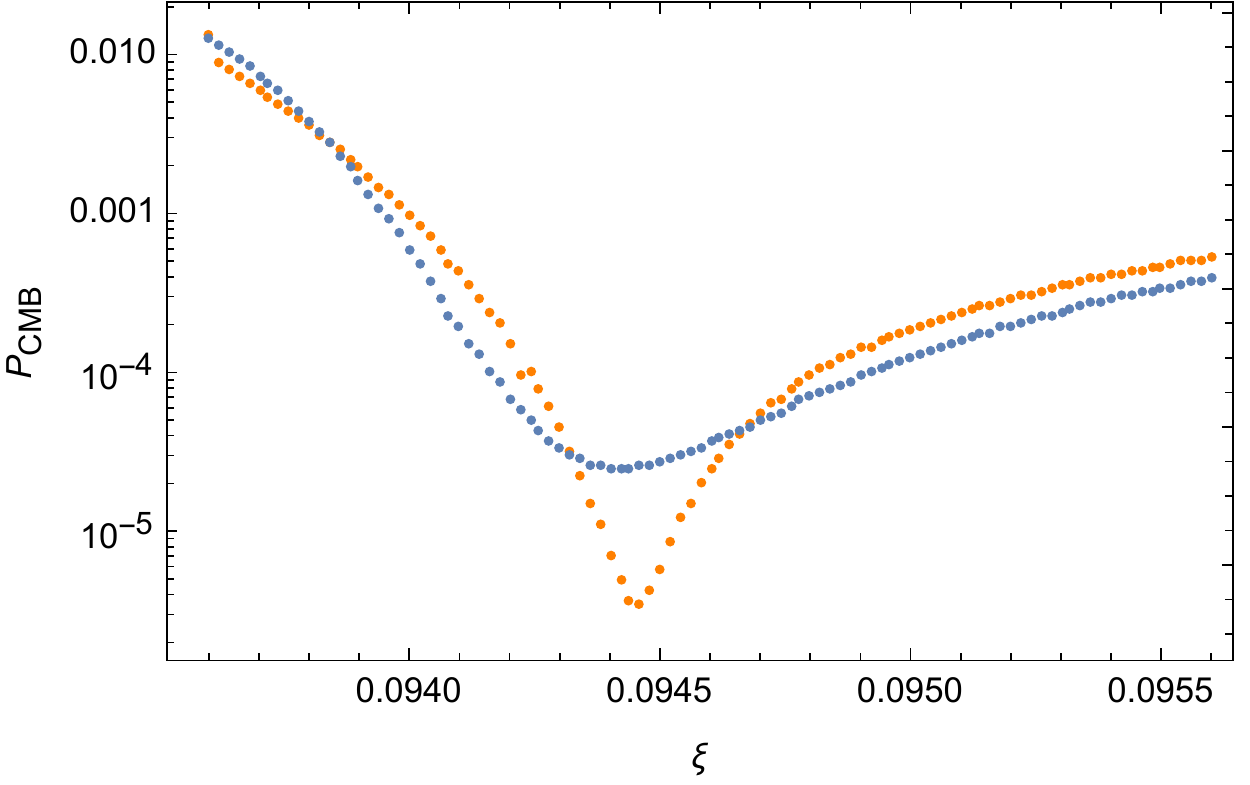}
		\caption{Scan over $\xi$ for finding the minimal CMB \\normalization}
		\label{Scan1a}
	\end{subfigure}
	\begin{subfigure}{.5\textwidth}
		\centering
		\includegraphics[width=.9\linewidth]{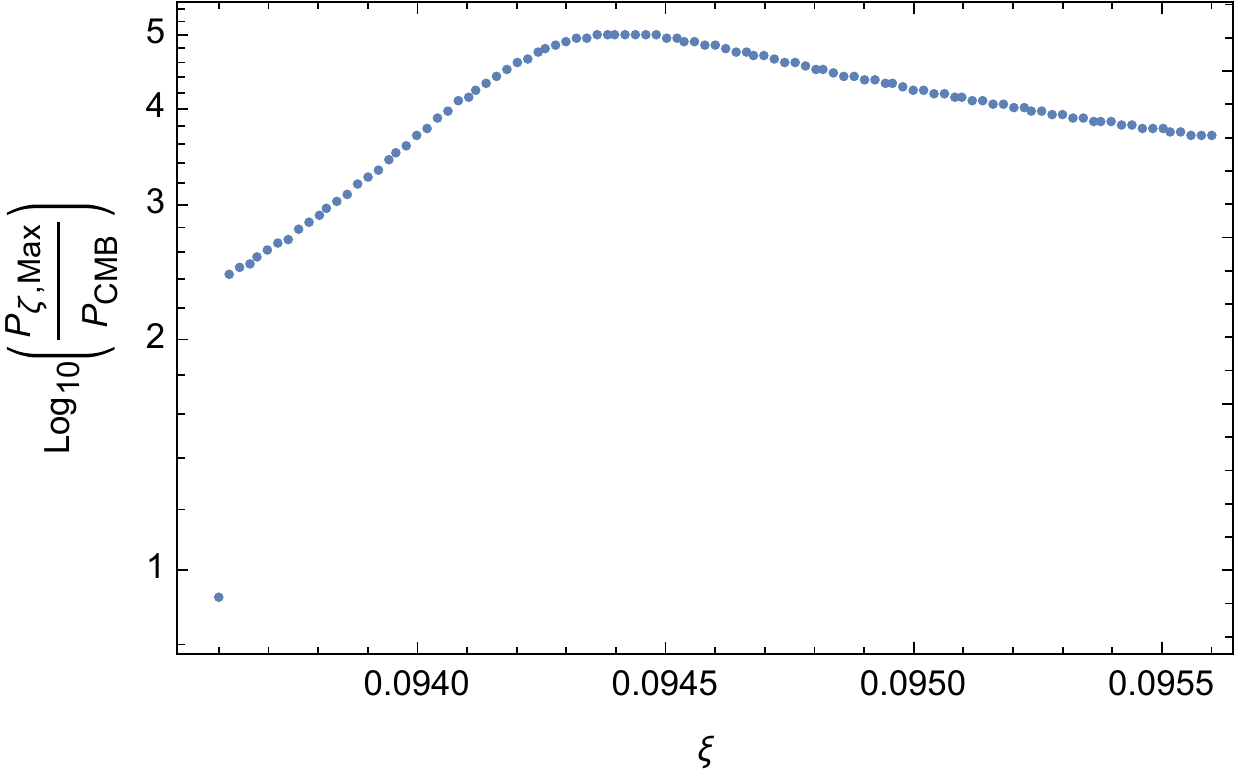}
		\caption{Scan over $\xi$ for finding the maximal \\enhancement of the power spectrum.}
	\end{subfigure}
		\begin{subfigure}{.5\textwidth}
		\centering
		\includegraphics[width=.9\linewidth]{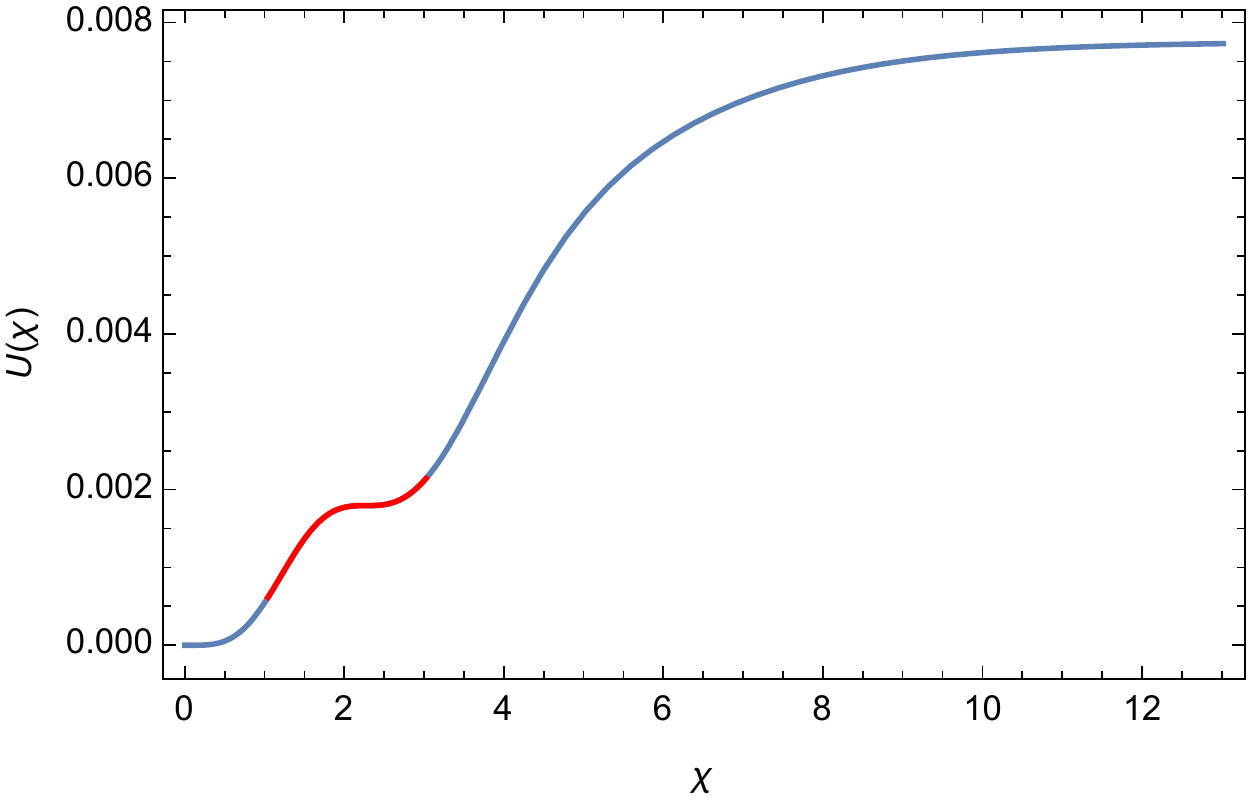}
		\caption{Inflationary dynamics for the parameters \\such that the CMB normalization is minimal.}
	\end{subfigure}
		\begin{subfigure}{.5\textwidth}
		\centering
		\includegraphics[width=.9\linewidth]{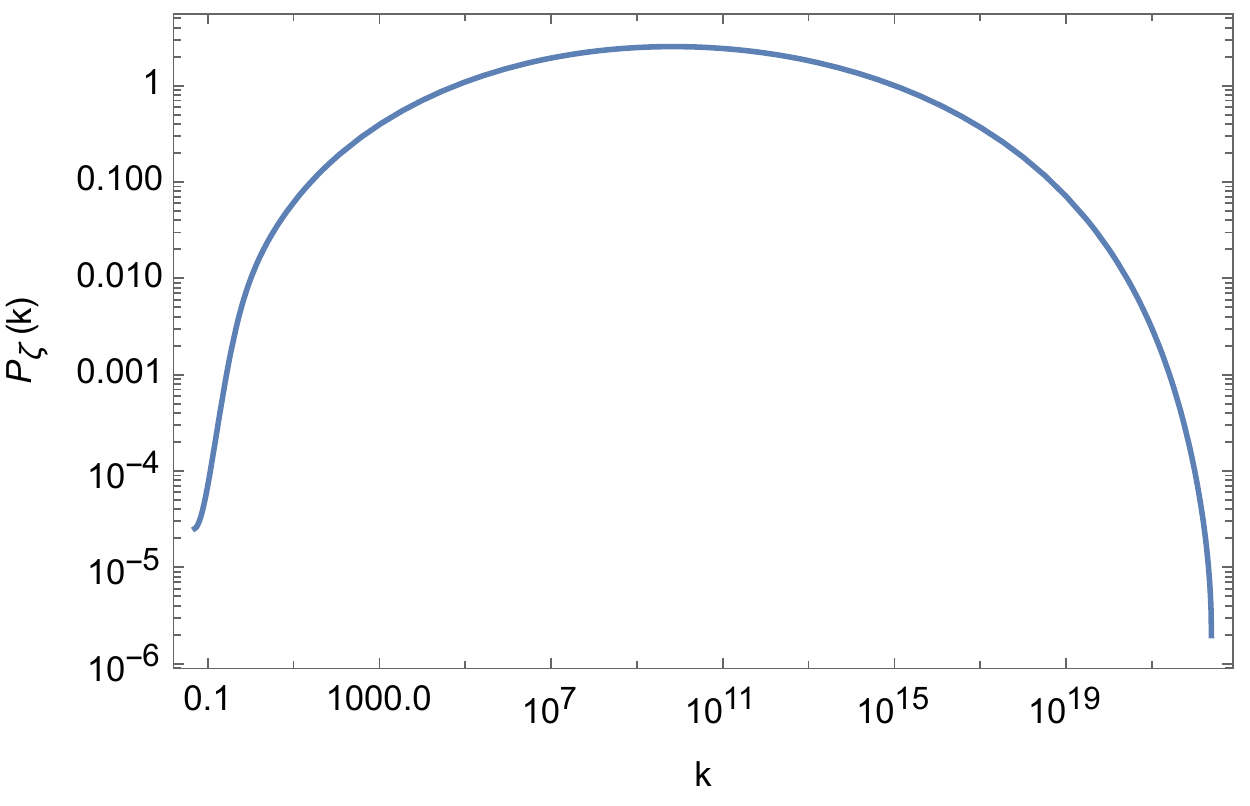}
		\caption{Power Spectrum as function of the scale for the parameters such that the CMB normalization is minimal.}
		\label{Scan1d}
	\end{subfigure}
	\caption{Scan Over $\xi$ for $y_t^{low} =0.919 $ and $\lambda_0 = 0.005$ which roughly corresponds to $m_t^{pole} = 170.25$ GeV. The \textbf{upper right} panel shows the value of the CMB normalization as function of $\xi$ around the minimal possible value where the values were obtained either by using equation \eqref{PowSR} (\textbf{in blue}) or equation \eqref{Mukhanov} (\textbf{in orange}). The \textbf{upper left} plot is showing the power spectrum enhancement as function of $\xi$. One can see that the minimal CMB corresponds to a peak in the enhancement.
	The two lower plots are representing the inflationary trajectory along the potential (\textbf{Lower Left}) and the power spectrum as function of the scale \textbf{Lower Right}.}
	\label{Scan1}
\end{figure}

\begin{figure}[H]
	\begin{subfigure}{.5\textwidth}
		\centering
		\includegraphics[width=.9\linewidth]{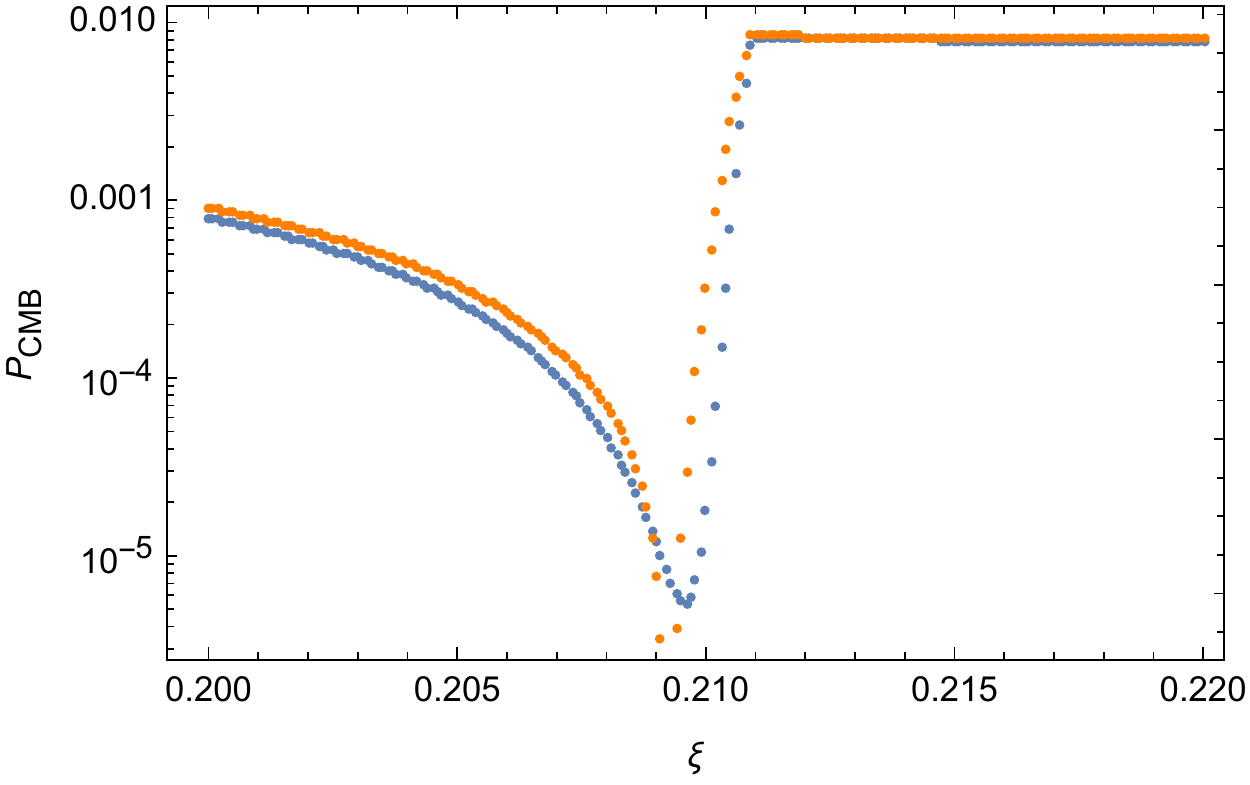}
		\caption{Scan over $\xi$ for finding the minimal CMB \\normalization}
		\label{Scan2a}
	\end{subfigure}
	\begin{subfigure}{.5\textwidth}
		\centering
		\includegraphics[width=.9\linewidth]{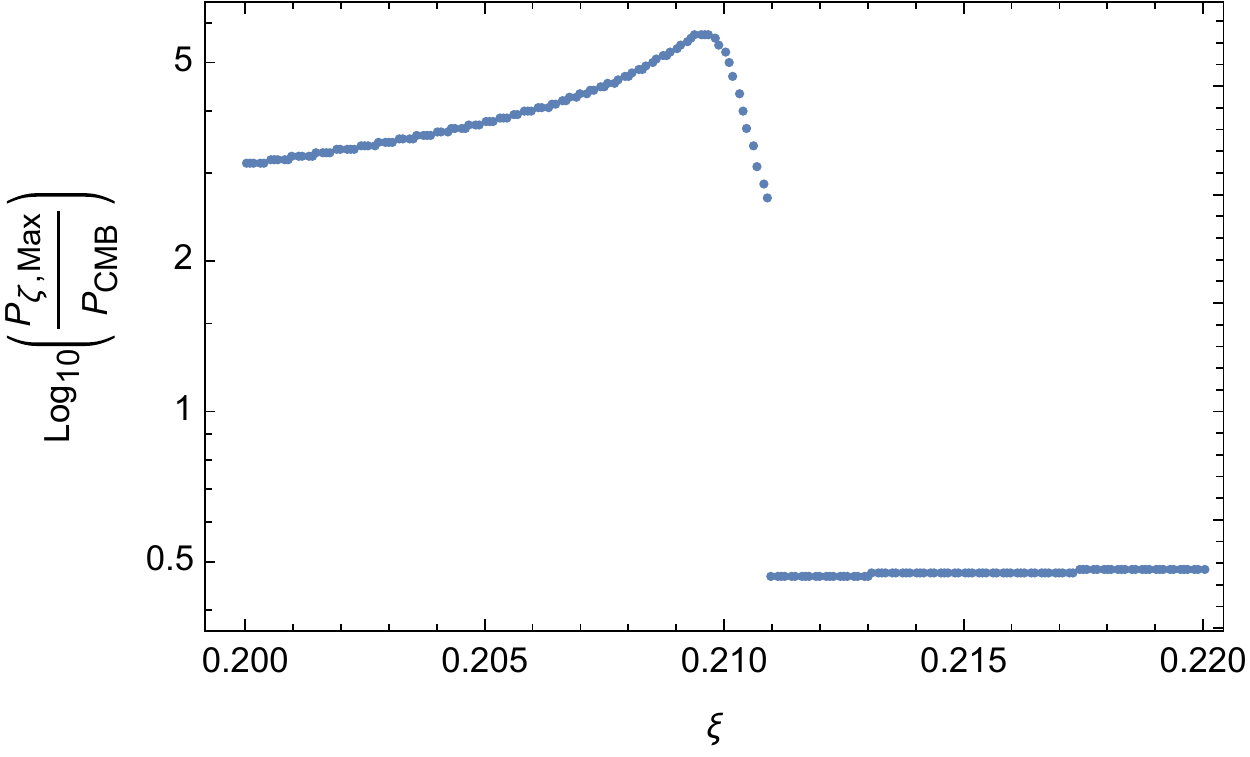}
		\caption{Scan over $\xi$ for finding the maximal \\enhancement of the power spectrum.}
	\end{subfigure}
	\begin{subfigure}{.5\textwidth}
		\centering
		\includegraphics[width=.9\linewidth]{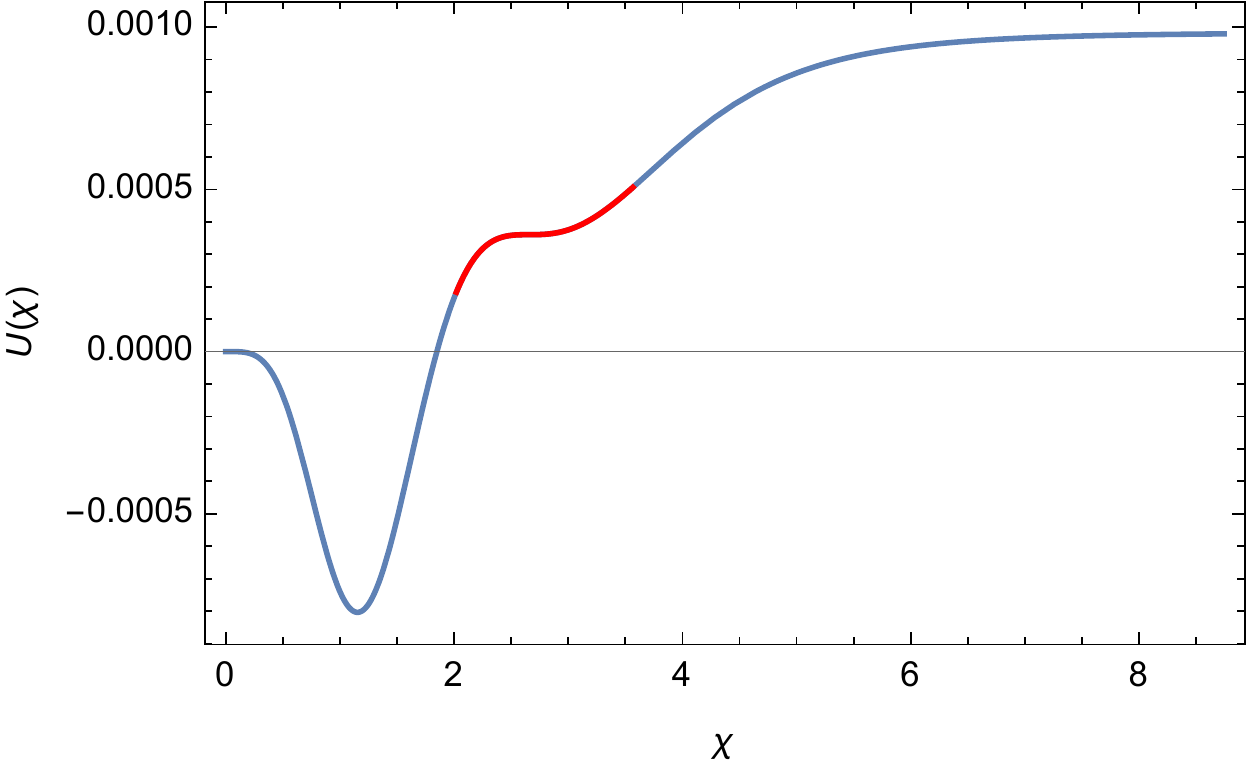}
		\caption{Inflationary dynamics for the parameters \\such that the CMB normalization is minimal.}
	\end{subfigure}
	\begin{subfigure}{.5\textwidth}
		\centering
		\includegraphics[width=.9\linewidth]{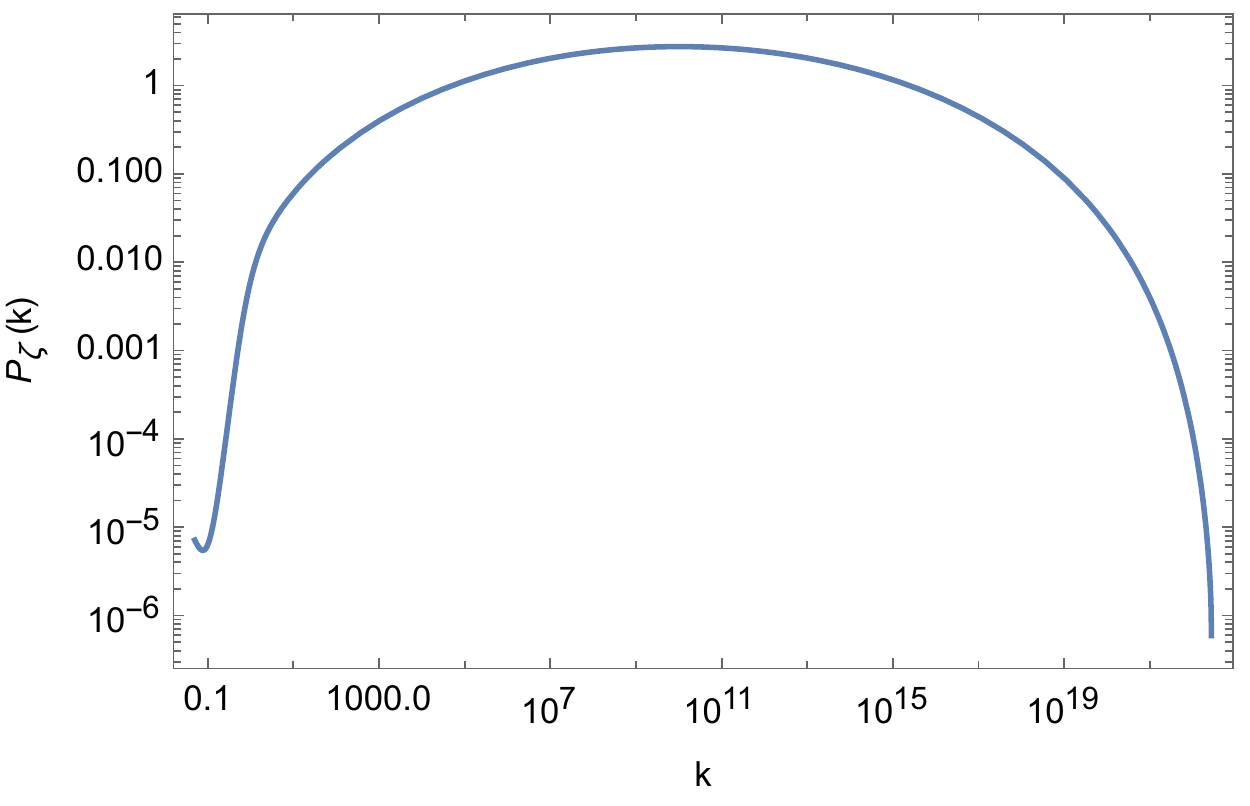}
		\caption{Power Spectrum as function of the scale for the parameters such that the CMB normalization is minimal.}
		\label{Scan2d}
	\end{subfigure}
	\caption{Scan over $\xi$ for $y_t^{low} =0.934 $ and $\lambda_0 = -0.012$ which roughly corresponds to $m_t^{pole} = 172.76$ GeV. The \textbf{upper right} panel shows the value of the CMB normalization as function of $\xi$ around the minimal possible value where the values were obtained either by using equation \eqref{PowSR} (\textbf{in blue}) or equation \eqref{Mukhanov} (\textbf{in orange}). The \textbf{upper left} plot shows the power spectrum enhancement as function of $\xi$. One can see that the minimal CMB corresponds to a peak in the enhancement. The two lower plots represent the inflationary trajectory along the potential (\textbf{Lower Left}) and the power spectrum as a function of the scale \textbf{Lower Right}.}
	\label{Scan2}
\end{figure}

The last thing one has to worry about before getting started is about the beginning of inflation. Sometime, our procedure selects some $\delta\lambda$ for which inflation is ending but where the CMB must be generated below the local maximum of the potential. To avoid this, we have to increase the value of the jump until the inflation correctly goes through an USR phase.
The results of this scan are pictured through the two example in figure \ref{Scan1} and \ref{Scan2}.

One can see from  figures~\ref{Scan1} and~\ref{Scan2} that contrary to the jumpless case, the system spends a long time in the USR phase. The reason for that is the presence of an additional free parameter which allows us to tune the depth of the local minimum such as to maximize the number of e-foldings spent within the local minimum. Contrary to what is done in figure \ref{ResultsJpLess}, we do not include the results of equation \eqref{Mukhanov} in the plots \ref{Scan1d} and \ref{Scan2d}. Since the system evolve mostly in USR, one cannot rely on the quadratic action \eqref{QuadActPert} and the numerical integration of \eqref{Mukhanov} results in $\mathcal{P}\gg \mathcal{O}(1)$. Thus, the only statement one can make about this parameter region is that we lose the perturbative control over the theory.

Figures \ref{Scan1a} and \ref{Scan2a} make evident that both of the cases are failing to reproduce the PLANCK constraints on the power spectrum (equation \eqref{PLANCK}). All the top Yukawa couplings we tested were characterized by a minimal CMB normalisation about three orders of magnitude above the required value.

One can also look at the behaviour of the spectral index and the tensor to scalar ratio. Since the SR violation comes from a USR phase where $\epsilon_H$ is even smaller than in the SR regime, we do not expect the tensor to scalar ratio to be inconsistent with \eqref{rPLANCK}. However, the spectral index strongly depends on $\eta_H$ and thus, it is not clear whether it will be consistent with \eqref{nsPLANCK}. In most parts of the parameter space, observational bounds are not respected.
\begin{figure}[H]
	\begin{subfigure}{.5\textwidth}
		\centering
		\includegraphics[width=.9\linewidth]{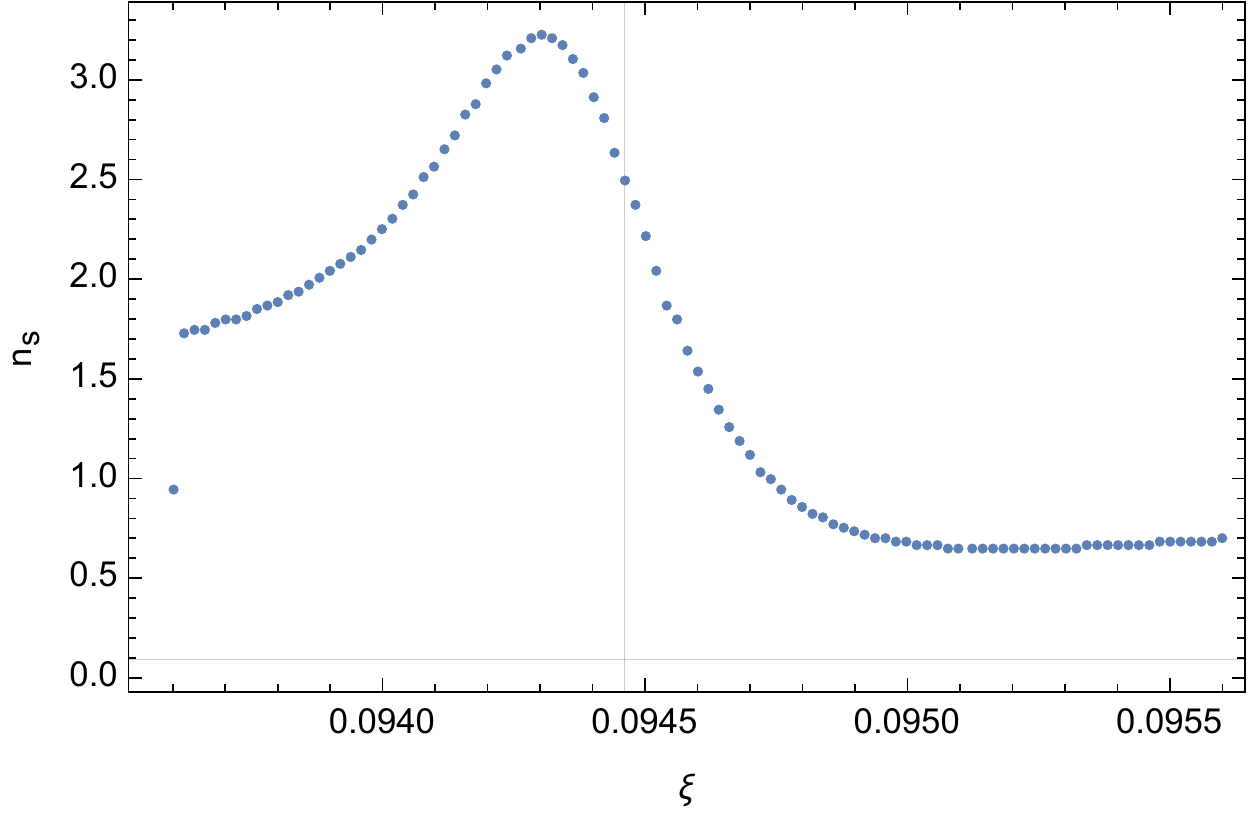}
		\caption{$n_s$ as function of $\xi$ for $y_t^{low} =0.919 $.}
	\end{subfigure}
		\begin{subfigure}{.5\textwidth}
		\centering
		\includegraphics[width=.9\linewidth]{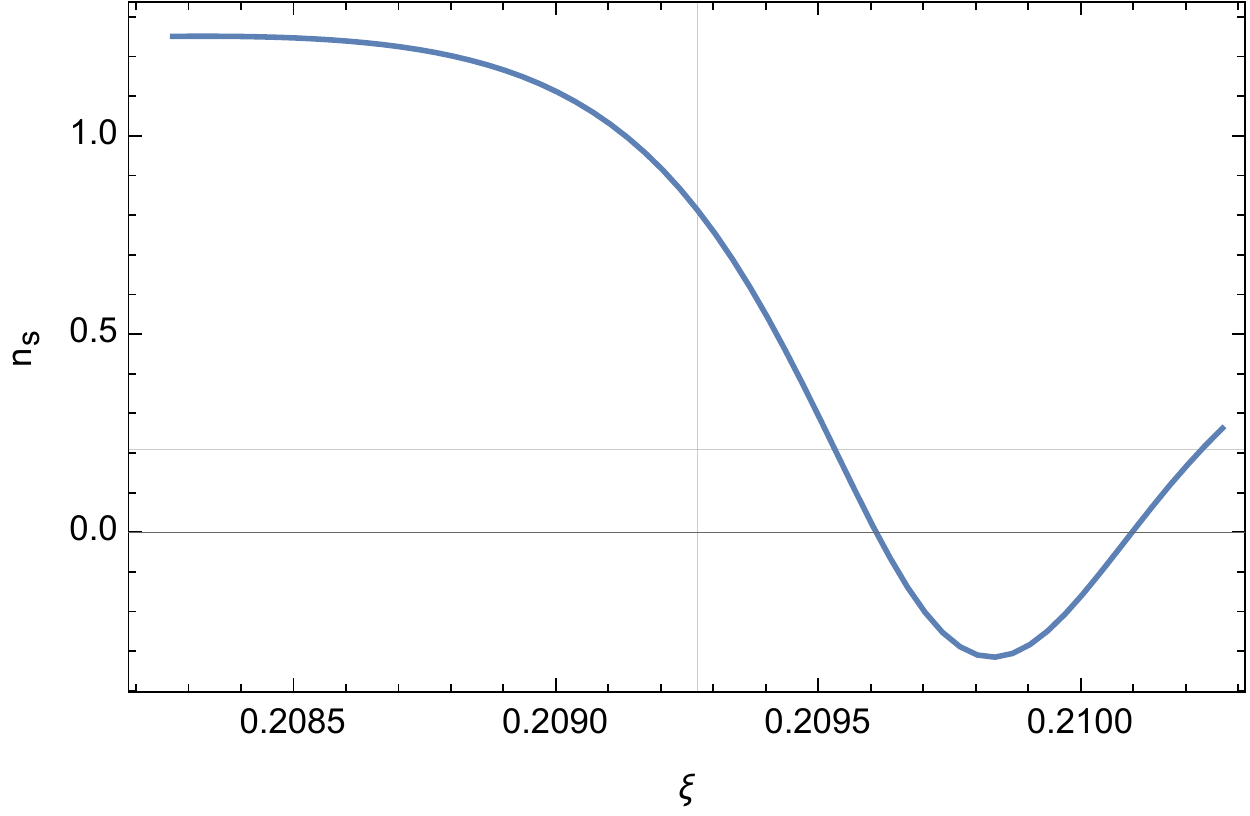}
		\caption{$n_s$ as function of $\xi$ for $y_t^{low} = 0.934$.}
	\end{subfigure}
	\caption{Representation of the spectral index as function of $\xi$ for the same top Yukawa as in figure \ref{Scan1} and \ref{Scan2}. The vertical lines indicate the locations of the minimal CMB.}
	\label{ResnsJump}
\end{figure}
\begin{figure}[H]
\begin{subfigure}{0.45\linewidth}
		\centering
		\includegraphics[width=\linewidth]{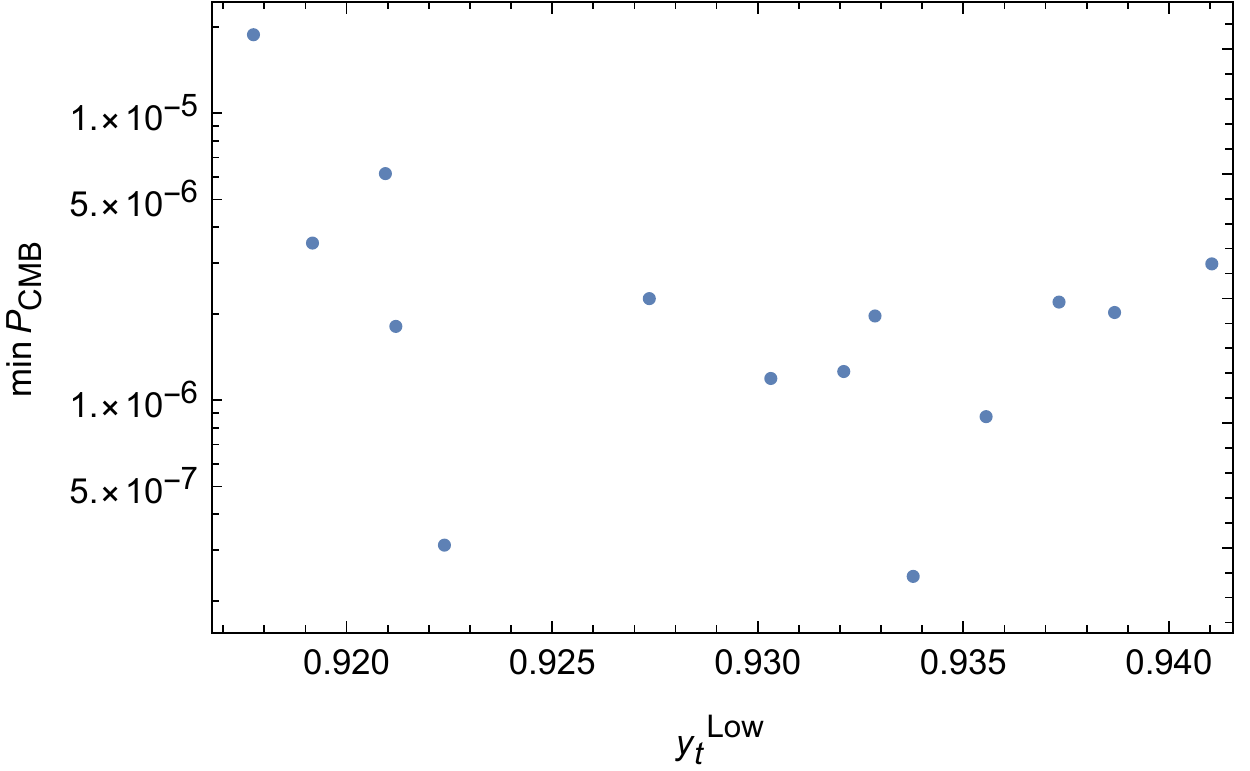}
		\caption{Results of the Scan procedure for several top Yukawas. The $y$ axis shows the minimal CMB normalization.}
		\label{MetaPlotCMB}
\end{subfigure}
\hspace{2em}
\begin{subfigure}{0.45\linewidth}
	\centering
	\includegraphics[width=\linewidth]{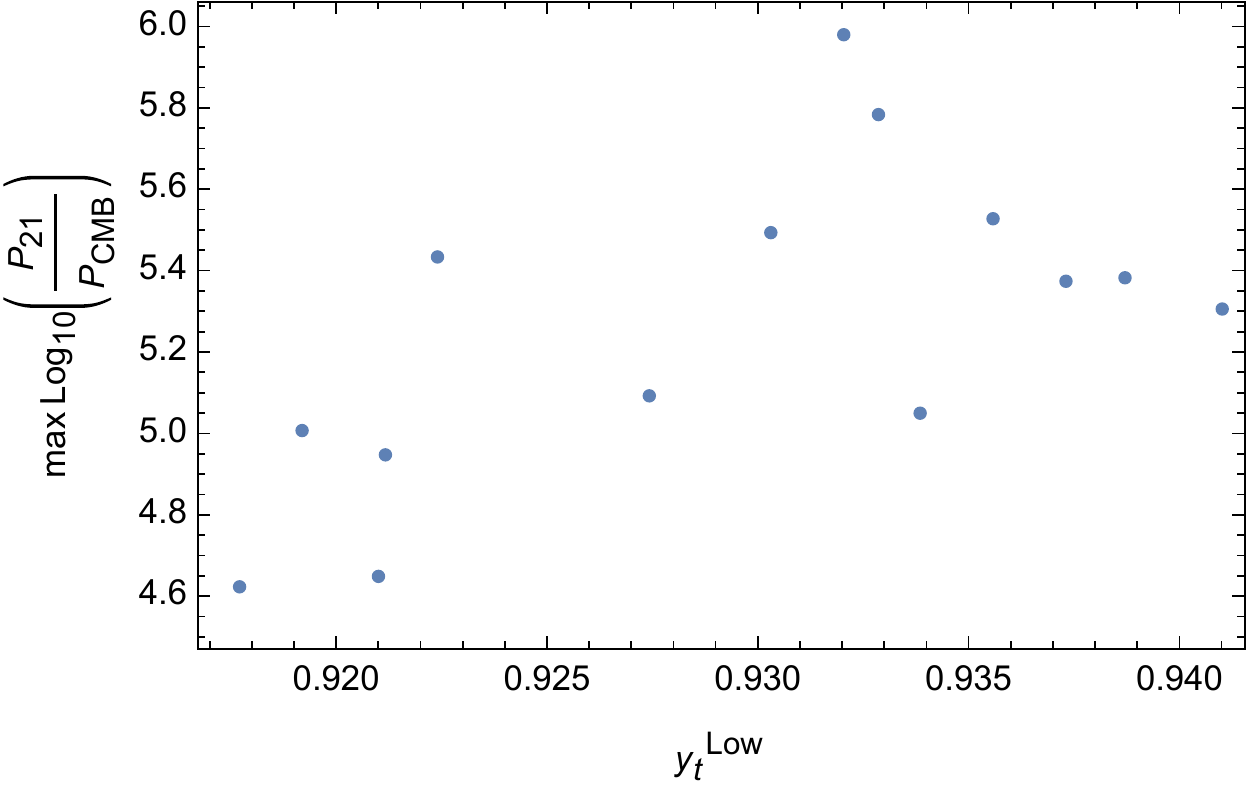}
	\caption{Results of the Scan procedure for several top Yukawas. The $y$ axis shows the maximal Enhancement.}
	\label{MetaPlotEnh}
\end{subfigure}
\caption{Result of all our scan at fixed $y_t^{low}$. Each time, we produced selected the minimal value for the CMB as well as the maximal value for the enhancement.}
\label{MetaPlot}
\end{figure}

Figure \ref{MetaPlot} summarises all the findings for the probed values of the top Yukawas. 
In all cases, we find the CMB normalization is greater than
$\mathcal{O}(10^{-7})$, which is at least two orders of magnitude above the observational constraint \eqref{PLANCK}.

\section{Conclusion} \label{sec:conclusion}
We have studied inflation driven by the Higgs field of the SM \cite{Bezrukov:2007ep,Bauer:2008zj}. This proposal is sensitive to the formulation of GR and we chose the Palatini version since it leads to an inflationary scenario with favourable properties at high energies. Namely, a comparatively high value of the cutoff scale \cite{Bauer:2010jg}, above which perturbation theory breaks down, contributes to a robustness against corrections originating from physics outside the SM; in particular, quantum corrections due to renormalization group (RG) running of coupling constants can become calculable with only small deviations \cite{Shaposhnikov:2020fdv}.

This motivated us to investigate parameter regimes in which RG running as computed within the SM can have large effects. Since the running Higgs self-coupling $\lambda(\mu)$ features a minimum slightly below the Planck mass $M_P$, this requires an inflationary scale $\mu$ on the same order. In Palatini Higgs inflation one has $\mu \sim M_P/\sqrt{\xi}$, where $\xi$ parametrises the strength of non-minimal coupling between the Higgs field and the gravitational Ricci scalar. As a result, we considered the regime $0<\xi \lesssim 1$. Additionally, we took into account two parameters that influence RG running. The first one is the low-energy value of the top Yukawa coupling -- varying it within its experimental uncertainty can have important effects. Finally, we use an effective jump of the Higgs self-interaction to parameterise possible contributions from unknown UV-physics. Consequently, our framework features three tunable coupling constants so that our initial expectation was to find viable inflationary scenarios, possibly with an associated production of primordial black holes due to a critical point in the potential.

However, our results were different. On the one hand, we concluded that a critical point corresponding to an inflection point or a small local minimum can arise as a result of RG running and lead to a large enhancement of the generated power spectrum. On the other hand, we found that the observed amplitude of perturbations in the CMB cannot be matched. 
Our results are fully consistent with previous findings, which showed that larger corrections to RG running can indeed give rise to critical points \cite{Rasanen:2018fom,Enckell:2020lvn}. So the difference of \cite{Rasanen:2018fom,Enckell:2020lvn} and our approach lies in assumptions about the magnitude of radiative effects. How drastically RG running can be altered at high energies is one of the questions about the UV-behaviour of Palatini Higgs inflation that remains to be answered.

In the end, we find our result about the absence of viable inflationary scenarios with a critical point -- in the regime $0<\xi\lesssim 1$ -- very interesting. This represents an expectation to the rule that a model of inflation with a sufficient number of parameter leads to a certain level of arbitrariness. Thus, one can view our finding as contributing to indications such as \cite{Bauer:2010jg,Rasanen:2017ivk,Markkanen:2017tun,Enckell:2018kkc,Shaposhnikov:2020fdv,Enckell:2020lvn} suggesting that Palatini Higgs inflation is robust against quantum corrections and exhibits some degree of universality in its predictions.

\section*{Acknowledgements}
 We thank Guillermo Ballesteros, Syksy R\"as\"anen, Mikhail Shaposhnikov, and Eemeli Tomberg for the helpful discussions and comments.
 
 The work of I.T. was partially supported by ERC-AdG-2015 grant 694896, by the Carlsberg foundation, and by
 the European Union's Horizon 2020 research and innovation program under the Marie Sklodowska-Curie grant agreement No. 847523 `INTERACTIONS'. S.Z. acknowledges support of the Fonds de la Recherche Scientifique - FNRS.

\appendix

\section{Small Non-Minimal Coupling Limit}
\label{app:SmallXi}
One of our main findings is that the generation of large perturbations requires a small non-minimal coupling $\xi<1$. With that in mind, it could be instructive to study what happens when one takes the limit $\xi\to 0$, where the function $F(\chi)$ reduces to
\begin{equation}
	\lim_{\xi\to 0}F(\chi) = \chi + \mathcal{O}\left(\xi\chi^3\right).
\end{equation}
Thus, as long as the Higgs field stays in the range where $\chi \ll 1/\xi^{1/2}$, the tree level potential of Palatini Higgs inflation reduces to the Jordan frame potential:
\begin{equation}
	U(\chi) = V(\chi) = \frac{\chi^4}{4}.
\end{equation}
This is expected since the same limit in the action \eqref{Act1} before the conformal transformation makes unnecessary the differentiation between these two frames as $\Omega\to 1$.\\
The full Palatini Higgs effective potential within the standard model quantum corrections is then given by:
\begin{equation}
	U(\chi) = \left(\lambda_0 + b \log^2\left(\frac{y_t\chi}{q}\right)\right)\frac{\chi^4}{4}.
\end{equation}
The slow roll parameter is given by:
\begin{equation}
	\epsilon_{SR}(\chi) = \frac{2}{\chi^2}\left(\frac{2\lambda_0 + b\log\left(\frac{y_t\chi}{q}\right) + 2 b\log^2\left(\frac{y_t\chi}{q}\right)}{\lambda_0 + b\log^2\left(\frac{y_t\chi}{q}\right)}\right)^2 \,,
\end{equation}
and the CMB normalization by:
\begin{equation}
	\frac{U(\chi)}{\epsilon_{SR}(\chi)} = \frac{\chi^6}{8}\frac{\left(\lambda_0 + b\log^2\left(\frac{y_t\chi}{q}\right)\right)^3}{\left(2\lambda_0 + b\log\left(\frac{y_t\chi}{q}\right) + 2 b\log^2\left(\frac{y_t\chi}{q}\right)\right)^2} \,.
	\label{NormPhi4}
\end{equation}
Since this is a bit complex to understand, let us look at two relevant limits.\\
First, it is possible to take the tree level theory by setting $b \to 0$. The slow roll parameter then reduces to:
\begin{equation}
	\epsilon_{SR}^{Tree} = \frac{8}{\chi^2}\implies \chi^{Tree}_{end} = 2\sqrt{2}\sim 2.8 \,.
\end{equation}
Then one can find the beginning of inflation as usual
\begin{equation}
	N_\star = \int^{\chi_i}_{\sqrt{8}}\frac{d\chi}{\sqrt{2\epsilon_{SR}(\chi)}}
	= \frac{\chi_i^2}{8}- 1 \implies \chi_i = \sqrt{8(N_\star + 1)}\sim 21.2 \,.
\end{equation}
Then, to satisfy the PLANCK normalization the coupling $\lambda_0$ has to respect:
\begin{equation}
	2.81\times 10^6 \lambda_0 = 5.0 \times 10^{-7}\implies \lambda_0 = 1.78\times 10^{-13} \,,
\end{equation}
in agreement with the well-known result of inflation with a quartic potential.
In the tree level regime where we neglect $b$ in front of $\lambda_0$, we found that $\lambda_0$ must be very small. This seems inconsistent with the actual order of magnitude of $b\sim 2\times 10^{-5}\gg \mathcal{O}(10^{-13})$. Thus, in order to consider the quantum behaviour, it is instructive to consider the opposite limit where the quartic coupling is negligible in front of $b$:
\begin{equation}
	\epsilon_{SR}(\chi) \sim \frac{2}{\chi^2}\left(2 + \frac{1}{\log\left(\frac{y_t \chi}{q}\right)}\right)^2 = 1 \iff \chi_{end}^{Loop}\sim 3.8 \,,
\end{equation}
where we solved this equation numerically. The number of e-folds is then given by:
\begin{equation}
	N_\star = \left[\frac{\chi^2}{8}-\frac{q^2}{8 e y_t^2}\text{Ei}\left(1+2\log\left(\frac{y_t \chi}{q}\right)\right)\right]^{\chi_i^{Loop}}_{3.8}\iff \chi_i^{Loop} = 23.06 \,.
\end{equation}
where we also solved the equation numerically.\\
This is neither a physical scenario since the free parameter of the model has no more influence on the CMB power spectrum. However, this tells us about the range where the boundaries of inflation can lie in:
\begin{equation}
	2.8\lesssim \chi_{end} \lesssim 3.8,\quad 21.2\lesssim\chi_i \lesssim 23.06 \,.
\end{equation}
This gives us the possible range of the initial $\log$:
\begin{equation}
	3.3\lesssim\log\left(\frac{y_t \chi_i}{q}\right)\lesssim 3.35 \,.
\end{equation}
Now, let us consider the equation \eqref{NormPhi4} by taking the center of the interval with $\chi_i = 22.13$ and $\log(y_t\chi_i)= 3.31$:
\begin{equation}
	\frac{U}{\epsilon}\sim 1.5 \times 10^7\frac{(11b + \lambda_0)^3}{(25.22 b + 2 \lambda_0)^2}\gtrsim \mathcal{O}(10^2) \,. 
\end{equation}
With such a form, this can never be small for such a large value of $b$. Indeed, one cannot take $\lambda_0$ large because of the cubic nature of the numerator and when $\lambda_0\to 0$, it can be neglected in front of the $b$ and we fall into the loop dominated regime which does not contains any free parameters to tune the initial power spectrum.\\
The formal limit of small non-minimal coupling of Palatini Higgs inflation is then improper for inflation once the quantum corrections from the standard model are involved.\\
The next thing to do is to take in account the action of the jump in the coupling constant. Actually, this does not improve the situation. Indeed, in the wanted limit, the function $G^2(\chi)$ is:
\begin{equation}
	\lim_{\sqrt{\xi}\chi\to 0}G^2(\chi) = 1 + \mathcal{O}(\chi^2\xi) \,.
\end{equation}
Since we already assumed $\chi\ll 1/\xi^{1/2}$, we can consider $G^2$ as a constant in that range.\\
Thus, the RG running's structure is exactly the same than in the jumpless case apart from a $\lambda_0\to\Bar{\lambda}_0$ replacement. In this situation, the only change is an increased freedom of the quartic coupling which is now promoted to a free parameter. However, this does not change anything to the argument above and the small $\xi$ limit is excluded as well. 
	
	\bibliographystyle{JHEP}
	\bibliography{PBHHI}
	
\end{document}